\documentclass[ 12pt]{article}
\usepackage{authblk}
\usepackage{ae}
\usepackage[T1]{fontenc}
\usepackage[ansinew]{inputenc}
\usepackage{amsmath}
\usepackage{amssymb}
\usepackage{color}
\usepackage{graphicx}
\usepackage{longtable}
\usepackage{latexsym}
\usepackage{stmaryrd}
\usepackage{amsfonts}
\usepackage{amsbsy}
\usepackage{mathrsfs}
\usepackage{psfrag}
\usepackage{enumerate}
\usepackage{MnSymbol}
\usepackage{vmargin}
\setmarginsrb{3cm}{2cm}{3cm}{2cm}{1cm}{1cm}{1cm}{1cm}
\usepackage{hyperref}
\usepackage{cite}

\usepackage{amsmath,amssymb,calc,amsfonts}
\usepackage{latexsym}

\usepackage{graphicx,calc,epsfig}

\newcommand{\C}{\mathbb{C}}
\newcommand{\R}{\mathbb{R}}
\newcommand{\n}{\nonumber}
\newcommand{\be}{\nopagebreak[3]\begin{equation}}
\newcommand{\ee}{\end{equation}}
\newcommand{\bee}{\nopagebreak[3]\begin{equation*}}
\newcommand{\eee}{\end{equation*}}
\newcommand{\ba}{\nopagebreak[3]\begin{eqnarray}}
\newcommand{\ea}{\end{eqnarray}}
\newcommand{\baa}{\nopagebreak[3]\begin{eqnarray*}}
\newcommand{\eaa}{\end{eqnarray*}}
\newcommand{\la}{\label}
\DeclareFontFamily{U}{rsfs}{}         
\DeclareFontShape{U}{rsfs}{m}{n}{<5> rsfs5 <6><7> rsfs7          %
  <8><9><10><10.95><12><14.4><17.28><20.74><24.88> rsfs10}{}     %
\DeclareMathAlphabet{\mathfs}{U}{rsfs}{m}{n}                     %
\newcommand{\mfs}[1]{\mathfs {#1}}                               %

\newcommand{\va}{\scriptscriptstyle}

\newcommand{\sH}{{\mfs H}}

\newcommand{\sM}{{\mfs M}}

\newcommand{\ssp}{ \langle P s,\;s^{\prime} \rangle}

\newcommand{\p}{ \langle s, \; s^{\prime} \rangle_{\va ph}}
\newcommand{\vani}{\scriptstyle}

\newcommand{\Ha}{{\sH}_{aux}}

\newcommand{\su}{\mathfrak{su}}

\newcommand{\tr}{\mathrm{tr}}

\newcommand{\half}{\frac{1}{2}}

\begin{document}

\title{Turaev-Viro amplitudes from 2+1\\ loop quantum gravity}

\author[1]{Daniele Pranzetti\thanks{daniele.pranzetti@gravity.fau.de}}
\affil[1]{{\it Institute for Quantum Gravity}

University of Erlangen-N\"urnberg (FAU), Germany}

\sloppy
\maketitle

\pagestyle{plain}

\begin{abstract}

The Turaev-Viro state sum model provides a covariant spin foam quantization of three-dimensional Riemannian gravity with a positive cosmological constant $\Lambda$. We complete the program to canonically quantize the theory in the BF formulation using the formalism of loop quantum gravity. In particular, we show first how quantum group structures arise from the requirement of the constraint algebra to be anomaly free. This allows us to generalize the construction of the physical scalar product, from the $\Lambda =0$ case, in the presence of a positive $\Lambda$. We prove the equivalence between the covariant and canonical quantizations by recovering the spin foam amplitudes.

\end{abstract}
\sloppy
\maketitle
\newpage{}

\section{INTRODUCTION}\la{Intro}

Three-dimensional  gravity  can be formulated as a Chern-Simons theory with gauge algebra given by the isometry algebra of the local solutions of Einstein equations \cite{Witten2}. The Riemannian theory with a positive cosmological constant $\Lambda$ in this formulation, corresponding to the compact gauge group $SU(2)$, was first quantized via the path integral technique in \cite{Witten}. A canonical quantization was later achieved via the so-called combinatorial quantization scheme \cite{Combinatorial}, unraveling very clearly the fundamental role played by the theory of quantum groups.

The realization of the equivalence between the covariant and canonical quantization of the Chern-Simons formulation of three-dimensional gravity led to deep and surprising relationships between quantum gravity and the theory of knot invariants. This was fully realized in \cite{RT}, where the connection with quantum groups was further emphasized.

BF theory provides an alternative formulation of three-dimensional Riemannian gravity, classically equivalent to the Chern-Simons one. In this formulation, the covariant quantization is performed via the spin foam approach \cite{spinfoams}. In the case of a vanishing cosmological constant this is given by the Ponzano-Regge model \cite{PR}. For a positive cosmological constant, the Turaev-Viro state sum \cite{TV} can be shown to provide a covariant quantization of the theory \cite{TV-grav} (see also \cite{Freidel} for the connection between Turaev-Viro model and gravity). Also in the BF formulation there exists an intimate relationship between positive cosmological constant and quantum groups. In fact, the Turaev-Viro model provides a regularization of the Ponzano-Regge state sum by replacing the Lie group $SU(2)$ [whose recoupling theory defines a partition function $Z(M)$ for a triangulated compact 3-manifold $M$] with its quantum deformation $U_q SL(2)$, where the deformation parameter $q$ is a root of unity. 

The connection at the quantum level between the Chern-Simons and BF formulations is then established by the correspondence $Z_{TV}(M)=|Z_{WRT}(M)|^2$  \cite{Walker, Turaev, Turaev2} between the Turaev-Viro state sum and the Witten-Reshetikhin-Turaev path integral with the Chern-Simons action for $SU(2)_k \otimes SU(2)_{-k}$, where $k$ is the level. See \cite{Alexandrov} for a review of all these relations. 

The canonical quantization of the BF formulation can be tackled using the loop quantum gravity (LQG) framework. The program for the $\Lambda=0$ case has been successfully completed in \cite{Noui-Perez}. Using ideas introduced in \cite{Reisenberger} to construct a projection operator via the path integral representation of the $\delta$-distribution, \cite{Noui-Perez} provided a well defined procedure to map the kinematical states into the kernel of the quantum curvature constraint implementing the dynamics of the theory. In this way, it was shown how the Ponzano-Regge amplitudes can be recovered from the physical scalar product between kinematical spin network states of the canonical loop quantization (see also \cite{Bonzom} for the connection between the canonical quantization and the symmetries of the Ponzano-Regge model\footnote{For the relation between the LQG program and the combinatorial quantization formalism approach to the quantization of three-dimensional gravity in the case of vanishing cosmological constant see \cite{Meusburger}.}).
 
The case of non-vanishing cosmological constant is less clearly understood due to the more complicated form of the curvature constraint and its associated algebra \cite{Anomaly}. A first attempt to circumvent these difficulties was proposed in \cite{Ansatz}. An alternative route was taken in \cite{crossing, crossing2}, which relied on rewriting the curvature constraint in terms of the holonomy of a non-commutative connection. By means of a preferred quantization map, it was shown that the crossing of the quantum non-commutative holonomies reproduces exactly Kauffman's $q$-deformed crossing identity, with $q$ the deformation parameter depending on $\Lambda$ \cite{KL}. The recovering of one of the two Kauffman brackets defining the Turaev-Viro model, using only structures of the standard $SU(2)$ kinematical Hilbert space, represents a promising starting point to obtain the Turaev-Viro amplitudes from the physical inner product of canonical LQG, in analogy with the $\Lambda=0$ case. Here we are going to complete this goal.

In Section \ref{sec:classical} we review the classical phase space of 2+1 gravity with $\Lambda\neq 0$. In Section \ref{sec:CQ} we briefly introduce the canonical quantization scheme of LQG. In Section \ref{sec:Lambda=0} we recall the definition of the physical scalar product in the case of vanishing cosmological constant. Section \ref{sec:Lambda>0} contains the original results of this work. After briefly recalling the quantization of the non-commutative holonomy performed in \cite{crossing} and the main elements of the combinatorial definition of the Turaev-Viro state sum model, we study the algebra of the curvature constraint written in terms of Wilson loops of the non-commutative connection. We show that the proper quantum analog of the classical algebra is not recovered unless the trace of the holonomy along an infinitesimal loop evaluates to the quantum dimension of the spin-$j$ representation coloring the loop. This condition is then used to construct the physical scalar product in terms of an appropriate projector operator, in analogy with the $\Lambda=0$ case. Relying on the chromatic evaluation, we finally prove that the Turaev-Viro amplitudes are recovered from such physical scalar product between kinematical spin network states. Hence the equivalence between the covariant and canonical LQG quantizations even in the case of a positive cosmological constant is shown, providing an important consistency check for the theory.

\section{CLASSICAL PHASE SPACE}\la{sec:classical}

We are interested in Riemannian three-dimensional gravity with cosmological
constant $\Lambda\ge 0$ in the first order formalism. The space-time $\sM$ is
a three-dimensional oriented smooth manifold and the action reads
\be\la{action}
S[e,\omega]=\int_{\sM}\tr[e\wedge
F(\omega)+\frac{\Lambda}{3} e \wedge e\wedge e]\,,
\ee
where $e$ is a $\su(2)$ Lie algebra valued
$1$-form, $F(\omega)$ is the curvature of the three-dimensional
connection $\omega$ and $\tr$ is a Killing form on
$\su(2)$. The space-time manifold has topology $\sM= \Sigma \times \R$, with $\Sigma $ a Riemann surface that for now we assume having arbitrary genus.

In order to perform the canonical analysis, we need to pull back the connection and the triad to the space-like surface $\Sigma$. By doing so, the configuration variable of the theory is represented by the two-dimensional connection
$A^i_a$ and its conjugate momentum is given by the electric field $E^b_j = \epsilon^{bc} e^k_c
\eta_{jk}$, where $a = 1, 2$ are space coordinate indices, $i, j
= 1, 2, 3$ are internal $\su(2)$ indices (raised and lowered by the Killing metric $\eta$) and $\epsilon^{ab}=-\epsilon^{ba}$ with $\epsilon^{12}=1$ (similarly $\epsilon_{ab}=-\epsilon_{ba}$ with $\epsilon_{12}=1$). The Poisson bracket among these variables is given  by
\be\la{symplectic}
\{A^i_a(x), E^b_j(y)\}=\delta^b_a \delta^i_j \delta^{(2)}(x,y)~.
\ee
Because of the underlying $SU(2)$ and diffeomorphism gauge invariance
the phase space variables are not independent and satisfy the following set of first class constraints. The
first one is the analog of the familiar Gauss law of Yang-Mills theory, namely
\begin{equation}\label{Gauss}
G_i\equiv D_a E^a_i=0~,
\end{equation}
\noindent where $D_a$ is the covariant derivative with
respect to the connection $A$. The constraint
\eqref{Gauss} encodes the condition that the connection be
torsionless and it generates infinitesimal $SU(2)$ gauge
transformation. The second constraint reads
\be\la{Curvature}
C^i=  \epsilon^{ab} F_{ab}^i(A)+\Lambda \epsilon_{cd}\epsilon^{ijk}E^c_j E^d_k=0\,.
\ee
This second set of first class constraints generate local ``translations.'' Diffeomorphisms invariance of three-dimensional gravity is associated to these two previous sets of constraints, i.e. diffeomorphisms can be written as linear combinations of the transformations generated by \eqref{Gauss} and \eqref{Curvature}.

In order to exhibit the underlying (infinite-dimensional) gauge symmetry Lie algebra it is convenient to smear  the constraints \eqref{Gauss} and \eqref{Curvature}  with arbitrary test fields $\alpha$ and $N$, which we take independent of the phase space variables; they read
\be\la{Gauss-sm}
G[\alpha]=\int_\Sigma \alpha^i G_i=\int_\Sigma \alpha^i D_a E^a_i=0
\ee
and
\be\la{Curvature-smeared} 
C_\Lambda[N]=\int_\Sigma N_i C^i=\int_\Sigma N_i(F^i(A)+\Lambda \epsilon^{ijk}E_j E_k)=0~.
\ee
The constraint algebra is then
\ba\label{algebra}
\{C_\Lambda[N],C_\Lambda[M]\}&=& \Lambda \ G[[N,M]]\n\\
\{G[\alpha],G[\beta]\}&=&G[[\alpha,\beta]]\n\\
\{C_\Lambda[N],G[\alpha]\}&=&C_\Lambda[[N,\alpha]],
\ea
where $[a,b]^i=\epsilon^{i}_{\ jk} a^jb^k$ is the commutator of $\su(2)$.

\section{CANONICAL QUANTIZATION}\la{sec:CQ}

The canonical quantization of the kinematics (i.e. the definition of the auxiliary Hilbert space where the constraints
are to be quantized) is well understood and we will just briefly review it here (see \cite{Noui-Perez} for more details).

Following Dirac's
quantization procedure one constructs first an auxiliary Hilbert
space $\Ha$ one which the phase space variables are represented.
 The key ingredient is the background-independent construction of this auxiliary Hilbert space.
The main input is to replace functionals of the connection by functionals of holonomies along paths (the so-called generalized connections) $\gamma \subset
\Sigma$: these are the basic excitations in terms of which the Hilbert space is constructed. The holonomy $h_{\gamma}[A]$ of the connection $A$ along a path $\gamma$ is given by
\be
\label{hol}h_{\gamma}[A]=P \exp\int_{\gamma} A \;.
\ee
The conjugate momentum (densitized triad) $E^a_i$ field is associated to its flux across codimension one surfaces. In the quantum theory then, holonomies and fluxes 
become operators acting on  $\Ha$ and the constraints have to be expressed in terms of these operators so that they satisfy the quantum analog of \eqref{algebra}.

The main ingredient to construct the auxiliary Hilbert space is represented by cylindrical functionals $\Psi_{\Gamma,f}[A]$, defined by a finite graph $\Gamma \subset \Sigma$ containing $N_\ell^{\va \Gamma}$ links and a
continuous function $f: SU(2)^{N_\ell^{\va\Gamma}}\rightarrow \C$ according to
\be \label{cyl}
\Psi_{\Gamma,f}[A]=f(h_{\gamma_1}[A],\dots,h_{\gamma_{N_{\ell}^{\va \Gamma}}}[A]),
\ee
where  the $\gamma_i$'s  label the links of $\Gamma$ on which the holonomies $h_{\gamma_i}[A]$ are defined. 
The scalar product between two cylindrical functions $\Psi_{\Gamma_1,f}[A]$ and $\Psi_{\Gamma_2,g}[A]$ is constructed by means of the  Ashtekar-Lewandowski measure \cite{AL} as
\ba
\label{innerk}\nonumber
\langle \Psi_{\Gamma_1,f},\Psi_{\Gamma_2,g}
\rangle &\equiv& \mu_{AL}(\overline{\Psi_{\Gamma_1,f}[A]}\Psi_{\Gamma_2,g}[A])\\ &=&\int\! \prod \limits_{i=1}^{N_\ell^{\va {\tilde \Gamma}}} dh_i \overline{\tilde f(h_{\gamma_1},\cdots,h_{\gamma_{N_\ell^{\va {\tilde \Gamma}}})}} \tilde g(h_{\gamma_1},\cdots,h_{\gamma_{N_\ell^{\va \tilde \Gamma}}})\,.
\ea 
In the previous expression $dh_i$ is the Haar measure on the $SU(2)$ group and the graph  $\tilde \Gamma \subset \Sigma$ is such that it includes both $\Gamma_1$ and $\Gamma_2$; $\tilde f, \tilde g$ denote the extensions of the functions $f, g$ to $\tilde \Gamma$, according to the prescription of \cite{AL}.  
The auxiliary Hilbert space $\Ha$ is  defined as the Cauchy completion of the space of cylindrical functionals $Cyl$
under (\ref{innerk}).

%

The (generalized) connection becomes a self-adjoint operator in the auxiliary Hilbert space represented by the quantum holonomy
\be
{\hat h_\gamma[A]} \Psi[A] \; = \; h_\gamma[A] \Psi[A]\,,
\label{ggcc}
\ee
acting by multiplication in $\Ha$.

The triad is associated with operators in $\Ha$ defining the flux of the electric field across one-dimensional lines. Namely, for a one-dimensional path $\eta^a(t)\in \Sigma$ we define
\be
E(\eta)\equiv \int E^{a}_i\tau^i n_a dt\,,
\ee
where $n_a\equiv
\epsilon_{ab}\frac{d\eta^a}{dt}$ is the normal to the path.
Therefore, the previous quantity represents the flux of $E$ across the
curve $\eta$.

The associated quantum operator in $\Ha$ can be defined from its action on holonomies. More precisely one has
\ba \label{flux}\hat E(\eta)\triangleright h_{\gamma}=\frac{1}{2}\hbar
\left\{\begin{array}{ccc} \!\! o(p) \tau_i h_{\gamma}\ \ \mbox{if $\gamma$
ends at $\eta$}\\ o(p) h_{\gamma} \tau_i\ \ \mbox{if $\gamma$ starts at
$\eta$}\end{array}\right., 
\ea
 where $o(p)=\pm1$ is the orientation of
the intersection $p\in \Sigma$ (denoted $p$ for puncture) of the pair of oriented curves in the order $(\eta,\gamma)$, namely \be
o(p)=\left.\frac{\epsilon_{ab}\dot{\eta}^{a}
\dot{\gamma}^{b}}{\left|\epsilon_{ab}\dot{\eta}^{a}
\dot{\gamma}^{b}\right|}\right|_{p}\, \ee at the intersection $p\in \Sigma$.  In other
words the operator $E(\eta)$ acts at a puncture as an $SU(2)$
left-invariant-vector field if the puncture is the source of $h_{\gamma}$,
and it acts as a right-invariant-vector field  if the puncture is the target
of $h_{\gamma}$.



%

\section{PHYSICAL SCALAR PRODUCT $(\Lambda=0)$}\la{sec:Lambda=0}

The quantization of three-dimensional Riemannian gravity with $\Lambda=0$ has been performed, in the LQG approach, in \cite{Noui-Perez}. In this case, by first introducing a regulator consisting of  an arbitrary finite cellular decomposition $\Delta_{\Sigma}$
of $\Sigma$---with plaquettes $p\in\Delta_{\Sigma}$ of coordinate area smaller or equal to $\epsilon^{2}$---the curvature constraint can be written as
\be\la{C0}
C_{0}[N]=\int_{\Sigma}\tr\left[N\,
F\left(A\right)\right]=\lim_{\epsilon\rightarrow0}\sum_{p\in\Delta_{\Sigma}}\tr\left[N_{p}\,
W_{p}\left(A\right)\right]\,, 
\ee
where $W_{p}(A)=1+\epsilon^{2}F(A)+o(\epsilon^{2})\in SU(2)$ is
the Wilson loop of the (commutative) Ashtekar-Barbero connection $A$ computed in the fundamental representation. The quantization
of the previous expression is straightforward as this Wilson loop acts
simply by multiplication on the kinematical states of 2+1 gravity, as described above. Then, the Ponzano-Regge amplitudes can be recovered through the definition of a physical scalar product by means of a projector operator into the kernel of \eqref{C0}.
More precisely, the physical inner product and the physical Hilbert space $H_{phys}$ of $2+1$ gravity with $\Lambda = 0$ can be defined starting from the formal expression for the generalized projection operator into the kernel of curvature constraint \cite{Reisenberger}:
\begin{equation}\label{eq:Projector}
P=``\prod_{x\in \Sigma}\delta(\hat{F}(A(x))"=\int
D[N]exp\left(i\int_\Sigma Tr [N \hat{F}(A)]\right)\,.
\end{equation}
In \cite{Noui-Perez} it has been shown how, introducing the regularization \eqref{C0} as an intermediate step for the quantization, this projector can be given a precise definition leading to a rigorous expression for the physical inner product of the theory which can be represented as a sum over spin foams whose amplitudes coincide with those of the Ponzano-Regge model. In details, given two spin networks $s$ and $s^{\prime}$ based on
the graphs $\Gamma$ and $\Gamma^{\prime}$, we order the set of
plaquettes $p\in \Delta_\Sigma^{\Gamma\Gamma^{\prime}}$ and define the
physical scalar product between $s$ and $s^{\prime}$ as
\ba\label{scalar}
 \p=\ssp &:=& \lim_{\epsilon\rightarrow 0} \ \ \langle\prod_{p} 
{\delta}(W_{p})s, \; s^{\prime} \rangle\n\\
&=& \lim_{\epsilon\rightarrow 0} \ \  \sum_{j_{\va p}}
 (2j_{\vani p}+1)  \langle\prod_{p}
 \chi_{j_{\va p}}({W}_{p})\ s, \;s^{\prime} \rangle,\n\\
 \ea
where the sum is over all half-integers $j_{\va p}$ labeling
each plaquette, ${W}_{p}$ is the holonomy around $p$ (based on
an arbitrary starting point) and $\chi_{j_{\va p}}({W}_{p})$
is the trace in the $j_{\va p}$ representation.

\section{PHYSICAL SCALAR PRODUCT $(\Lambda>0)$}\la{sec:Lambda>0}

The canonical treatment of the quantization of 2+1 gravity recalled in the previous section sets the bases for the extension of the analysis to the non-vanishing cosmological constant case. More precisely, we observe that, if we replace $W_{p}(A)$ in \eqref{C0} by $W_{p}(A_{\va \Lambda}^{\pm})$, with
\be\la{conn}
A_{\va \Lambda}^{\pm}=A\pm\sqrt{\Lambda} e
\ee
corresponding to the  self-dual and anti-self-dual components of the Spin(4) connection\footnote{The (classical) equivalence between the BF and Chern-Simons formulations of three-dimensional gravity is encoded in the fact that the local isometry algebra and the local gauge symmetry coincide.} associated to the Chern-Simons formulation of three-dimensional Riemannian gravity with positive cosmological constant, it is easy to see that at the classical level we get
\be\la{Cq}
C_{\Lambda}\left[N\right]=\lim_{\epsilon\rightarrow0}\sum_{p\in\Delta_{\Sigma}}\tr\left[N_{p}\,
W_{p}\left(A_{\va \Lambda}^{\pm}\right)\right]\mp G\left[\sqrt{\Lambda}N\right]\,.
\ee
 This provides a candidate background-independent regularization of
the curvature constraint $C_{\Lambda}[N]$ for arbitrary values of
the cosmological constant. Notice that on gauge invariant states
(i.e. the solution space of the Gauss constraint) the second term
simply drops out and it is
enough to impose that only one of the connections $A_{\va \Lambda}^{\pm}$ be flat in order to project onto the
physical Hilbert space.

The first steps towards the quantum imposition of the constraints \eqref{Cq} have been made in \cite{crossing}, where the quantization of the holonomy of the general non-commutative connection $A_{\lambda}\equiv A+\lambda e$ (for $\lambda\in \R$) has been defined. This analysis showed how, by means of a preferred quantization prescription induced by the Duflo map \cite{Duflo} in order to deal with ordering ambiguities, the crossing between quantum holonomies reproduces exactly Kauffman's $q$-deformed crossing identity\footnote{The idea to use the Duflo map to solve ordering ambiguities related to the non-commutativity of the operators associated to the triad field in the quantum theory \cite{zapa} was originally proposed in \cite{HT}. Also in this framework the application of the Duflo map led to the appearance of quantum group structures. See also \cite{Majid, Guedes} for a more mathematical analysis of the Duflo map role in the context of $2+1$ quantum gravity.}. More precisely, given two crossing paths $\eta$ and $\gamma$, the action of the quantum holonomy $h_{\eta}(A_{\lambda})$ on
$h_{\gamma}(A_{\lambda})$ can be expressed as
\be\la{cross1}
 h_{\eta}\left(A_{\lambda}\right)\triangleright
h_{\gamma}\left(A_{\lambda}\right)\,\left|0\right\rangle
=\begin{array}{c}
\includegraphics[width=1.4cm]{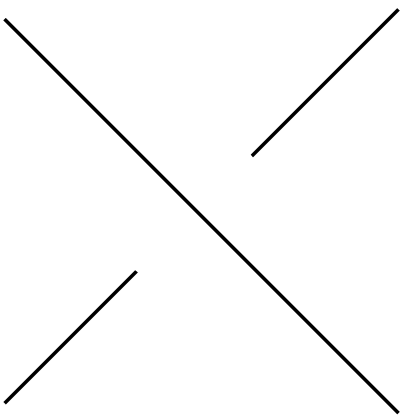}\end{array}=A  \begin{array}{c}
\includegraphics{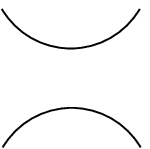}\end{array}+A^{-1}\begin{array}{c}
\includegraphics{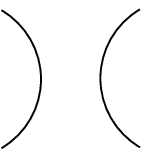}\end{array}
\ee
where $A=e^{\frac{i o\hbar \lambda}{4} }$, with $o$ the relative orientation between $\eta$ and $\gamma$. In this way,
the three-dimensional structure depicted as overcrossing or undercrossing encodes operator ordering and has to be understood as the link ``above'' (in this case $\eta$) acting on the link ``below'' (in this case $\gamma$). Analogously, 
\be\la{cross2}
 h_{\gamma}\left(A_{\lambda}\right)\triangleright
h_{\eta}\left(A_{\lambda}\right)\,\left|0\right\rangle
=\begin{array}{c}\psfrag{a}{}\psfrag{b}{}
\includegraphics[width=1.4cm]{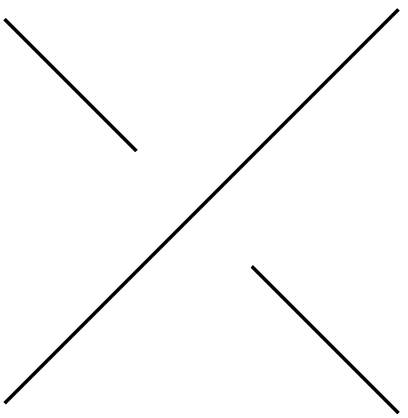}\end{array}=A^{-1}  \begin{array}{c}
\includegraphics{qL2.eps}\end{array}+A\begin{array}{c}
\includegraphics{qL1.eps}\end{array}\,.
\ee

The expressions \eqref{cross1} and \eqref{cross2}
have the same form as Kauffman's $q$-deformed binor identities
for $q=A^2=e^{\frac{i \hbar \lambda}{2} }$ (see below).

In the following, we are going to use this result in order to construct a projector operator into the kernel of the quantum version of the constraints \eqref{Cq}.
To this end, it will be useful to recall the basic notions of the Temperley-Lieb algebra; this is done in the next subsection. In \ref{sec:Algebra} we study the algebra of the quantum vesion of the constraint \eqref{Cq}.
This will be used in section \ref{sec:Projector} to construct a projector operator entering the definition of the physical scalar product, in analogy with the $\Lambda=0$ case, and hence compute physical transition amplitudes.

\subsection{Temperley-Lieb algebra and recoupling theory}\la{sec:Temperley-Lieb}

Kauffman's bracket polynomial \cite{KL} provides a tangle-theoretic interpretation of the Temperley-Lieb algebra and a combinatorial approach to the construction of 3-manifold topological invariants, such as the Turaev-Viro state sum model. Let us recall its definition and elementary properties. This section follows the presentation of \cite{KL}.

Given an unoriented link diagram $K$, a state $S$ of $K$ is a choice of smoothing for each crossing in $K$, where for the smoothing there are two possibilities labeled by $A, A^{-1}\in \C$. Thus $S$ appears as a disjoint set of Jordan curves in the plane decorated with labels at the site of each smoothing. 

Given a state $S$ of a diagram $K$, we denote by $||S||$ the number of disjoint Jordan curves in $S$ and by $\langle K|S\rangle$ the product of the state labels of $S$. With this notation, the {\it bracket polynomial} $\langle K\rangle$ is defined as the state summation 
\be\la{bracket}
\langle K\rangle = \sum_S \langle K|S\rangle d^{||S||},
\ee
where $S$ runs over all states (smoothing possibilities) of the diagram $K$, and $d= -A^2-A^{-2}$. The bracket polynomial satisfies the following properties:
\begin{enumerate}[(i)]
\item \bee \left< \begin{array}{c}\psfrag{a}{}\psfrag{b}{}
\includegraphics[width=0.8cm]{x-quantum.eps}\end{array}\right>=A\left<\begin{array}{c}
\includegraphics[width=0.8cm]{qL2.eps}\end{array}\right>+
A^{-1} \left<\begin{array}{c}
\includegraphics[width=0.8cm]{qL1.eps}\end{array}\right>,\la{Kau1}\eee
where all the diagrams stand for parts of larger ones, differing only at the given crossing. For the other type of crossing one has
\bee
\left< \begin{array}{c}\psfrag{a}{}\psfrag{b}{}
\includegraphics[width=0.8cm]{x-quantum-2.eps}\end{array}\right>=A^{-1} \left<\begin{array}{c}
\includegraphics[width=0.8cm]{qL2.eps}\end{array}\right>+
A\left<\begin{array}{c}
\includegraphics[width=0.8cm]{qL1.eps}\end{array}\right>.
\eee

\item \bee \left< \begin{array}{c}\psfrag{a}{}\psfrag{b}{}
\includegraphics[width=0.6cm]{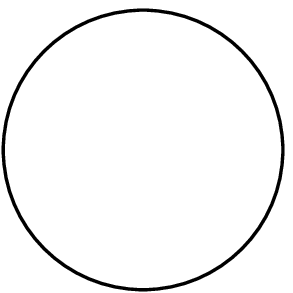}\end{array}\sqcup K\right>=d\left<K\right>,\la{Kau2}\eee
where $\begin{array}{c}\psfrag{a}{}\psfrag{b}{}
\includegraphics[width=0.4cm]{Loop.eps}\end{array}\sqcup$ denotes disjoint union of the diagram $K$ with a loop curve in the plane.
\end{enumerate}

Properties (i) and (ii) are called {\it Kauffman brackets} and they guarantee that the bracket polynomial is an invariant of regular isotopy of link diagrams; i.e. it satisfies the Reidemeister moves of type II and III (plus underlying graphical changes induced by homeomorphisms of the plane). 

Kauffman brackets were at first introduced as a deformation of Penrose spin network theory \cite{Penrose}; only in a second moment this recoupling theory was shown to be strictly related to the theory of $q$-deformed angular momentum recoupling using the quantum group $U_q(su(2))$ \cite{KL, q-def}.

Let us now consider tangles with $n$ incoming and $n$ outgoing strands (see, e.g., the figure below). The linear combinations of these $n$-tangles form the, so-called, {\it Temperley-Lieb algebra} $TL_n$, where, given two $n$-tangles $T$ and $S$,  the multiplication is defined by joining the outgoing strands of the one with incoming of the other vertically, namely
\begin{figure}[h]
\centerline{\hspace{0.5cm} \(
\begin{array}{c}
\includegraphics[width=5cm]{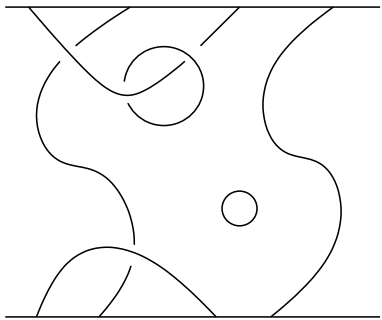}
\end{array}
\) }
\caption{Example of an $n$-tangle with $n=4$ incoming and outgoing strands.} \label{Tangle}
\end{figure}
\be
\begin{array}{c}
\includegraphics[width=1.5cm]{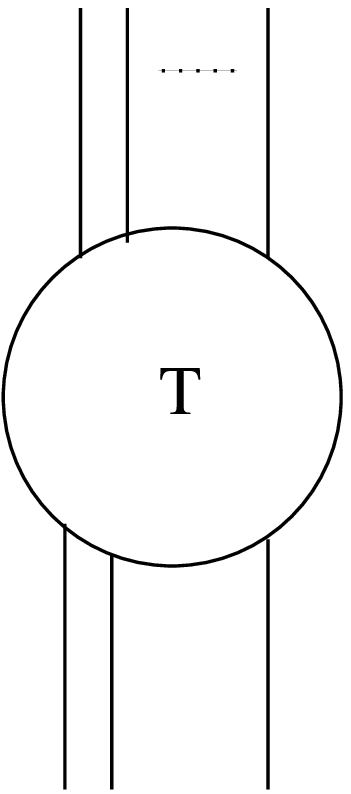}\end{array}\ast  \begin{array}{c}
\includegraphics[width=1.5cm]{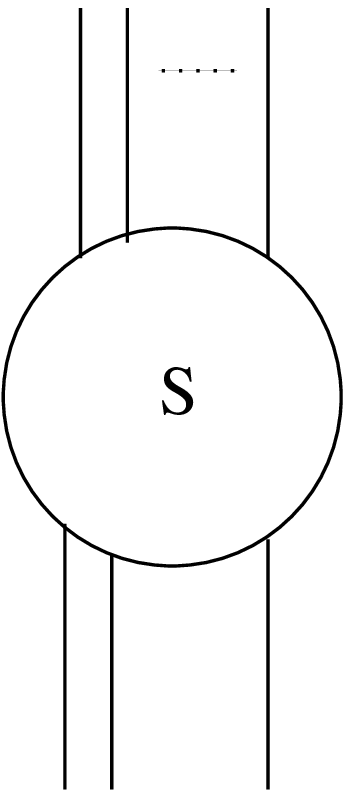}\end{array}=\begin{array}{c}
\includegraphics[width=1.5cm]{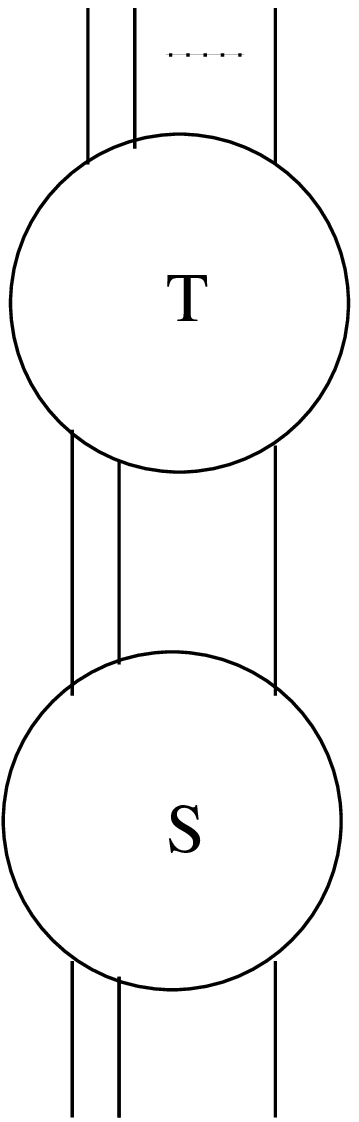}\end{array}=\begin{array}{c}
\includegraphics[width=1.5cm]{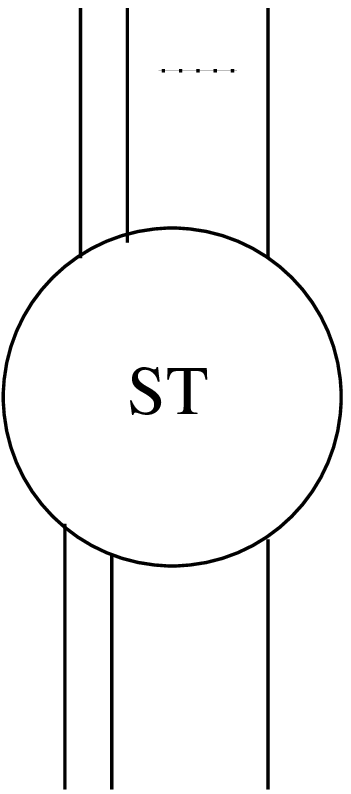}\end{array}
\label{mult}.\ee
By means of the skein relations (\ref{Kau1}) and (\ref{Kau2}), any element of $TL_n$ can be written as a linear combination of products of the {\it elementary tangles} $1_n, e_1, e_2\dots e_{n-1}\in TL_n$, where $1_n$ is the $n$-tangle that connects the $i$th input with the $i$th output and
\be
e_1= \begin{array}{c}\includegraphics[width=1.8cm]{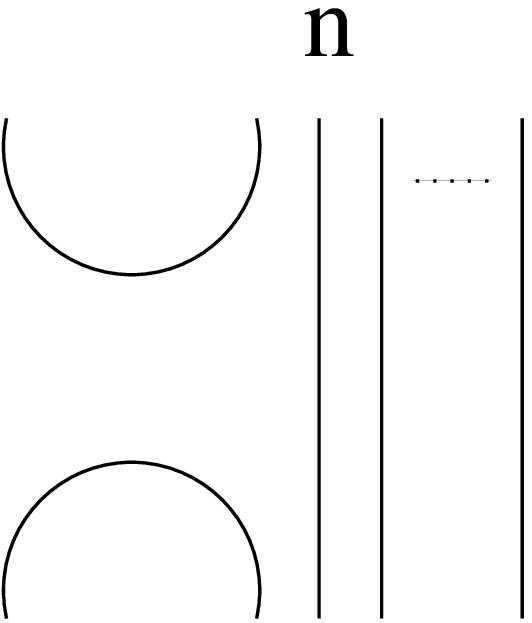}\end{array}~~~
e_2=\begin{array}{c}\includegraphics[width=1.8cm]{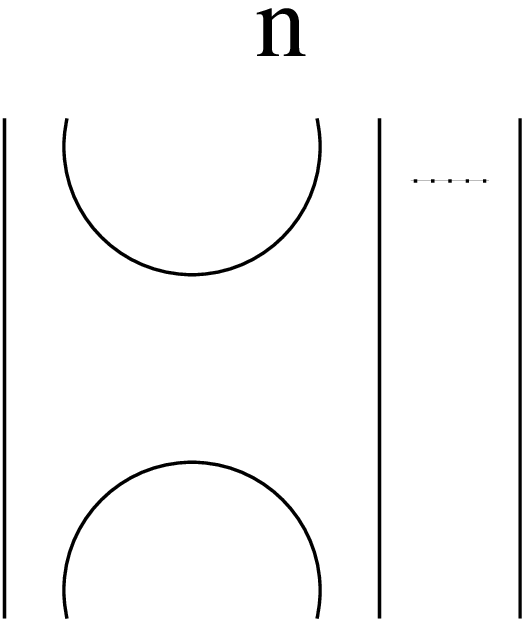}\end{array}\cdots~~
e_{n-1}=\begin{array}{c}\includegraphics[width=1.8cm]{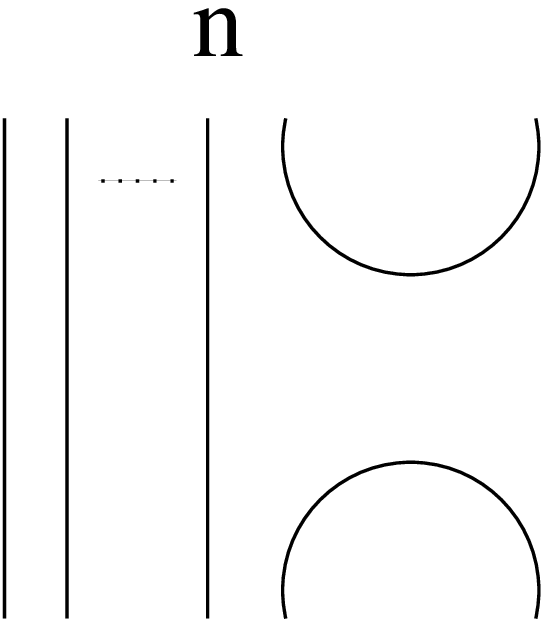}\end{array}
.\ee
The $e_i$'$s$ satisfy the following relations 
\ba
&&e_i^2=de_i\,;\la{e1}\\
&&e_ie_{i+1}e_i=e_i~~~{\rm and}~~~e_{i+1}e_ie_{i+1}=e_{i+1}\,;\la{e2}\\
&&e_ie_j=e_je_i~~~{\rm if}~|i-j|\leq 2\,.\la{e3}
\ea
Two products of elementary tangles represent equivalent tangles; i.e. they are regularly isotopic relative to their end points, if and only if one product can be obtained from the other by the relations (\ref{e1})--(\ref{e3}) above.
The Temperley-Lieb algebra $TL_n$ is the algebra freely generated by the multiplicative generators $1_n, e_1, e_2\dots e_{n-1}$ modulo (\ref{e1})--(\ref{e3}).

If $T$ is an $n$-tangle, let $\bar T$ denote the {\it standard closure} of $T$ obtained by attaching the $i$th input to the $i$th output; in a graphical notation
\be
T=\begin{array}{c}\includegraphics[width=1.3cm]{Tangle.eps}\end{array}~~~~~~~~
\bar T=\begin{array}{c}\includegraphics[width=3.2cm]{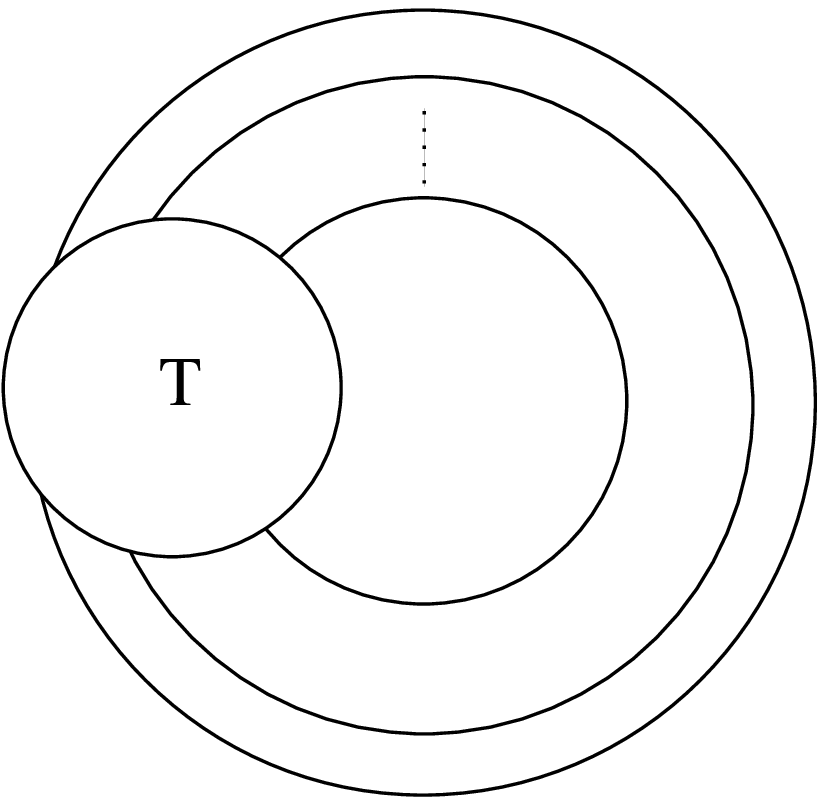}\end{array}.
\ee
We now define the {\it trace} of an $n$-tangle $T$ as $tr(T)=\langle\bar T\rangle$, where $\langle \rangle$ denotes the bracket polynomial (\ref{bracket}), and $tr(T+S)=tr(T)+tr(S)$. Note that $tr(TS)=tr(ST)$, as an immediate consequence of the properties of the bracket polynomial and of the form of the closure for the tangles. If $T\in TL_n$ is a product of $e_i$'$s$, then $\bar T$ is a disjoint union of Jordan curves in the plane, and $tr(T)=\langle\bar T\rangle= d^{||\bar T||}$, where $||\bar T||$ denotes the number of Jordan curves.

Let us now introduce the Artin group $B_n$ and its representation to the Temperley-Lieb algebra. Elements of $B_n$ are a special case of $n$-tangles, more precisely, a {\it braid} $b\in B_n$ is an $n$-tangle that is regularly isotopic to a product of {\it elementary braids} $1_n, \sigma_1,\dots,\sigma_{n-1}, \sigma^{-1}_1,\dots,\sigma_{n-1}^{-1}$, where the elementary braid $\sigma_i^\pm$ takes the input $i$ to the output $i+1$ and the input $i+1$ to the output $i$. The braids $\sigma_i$ and $\sigma_i^{-1}$ have opposite crossing type, namely
\be
\sigma_i=\begin{array}{c}\includegraphics[width=3cm]{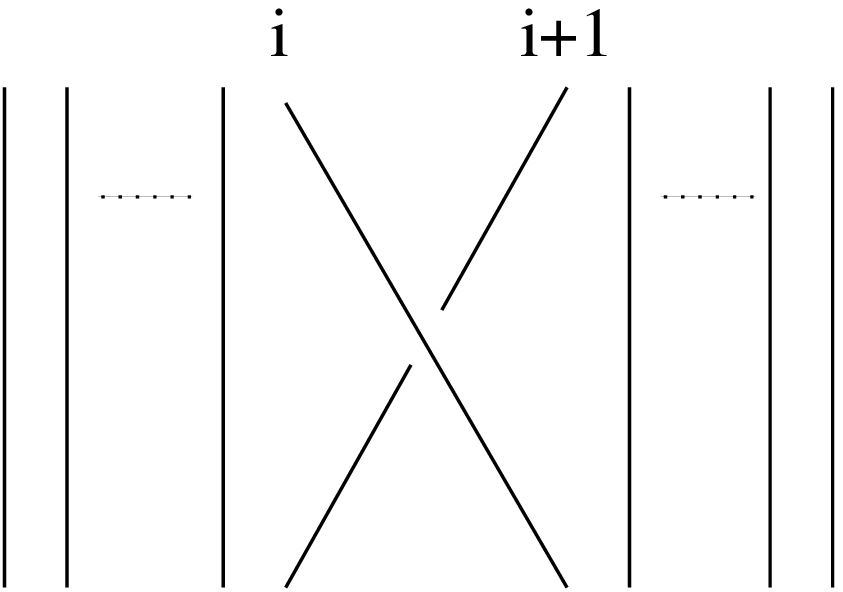}\end{array}~~~~~~~~
\sigma_i^{-1}=\begin{array}{c}\includegraphics[width=3cm]{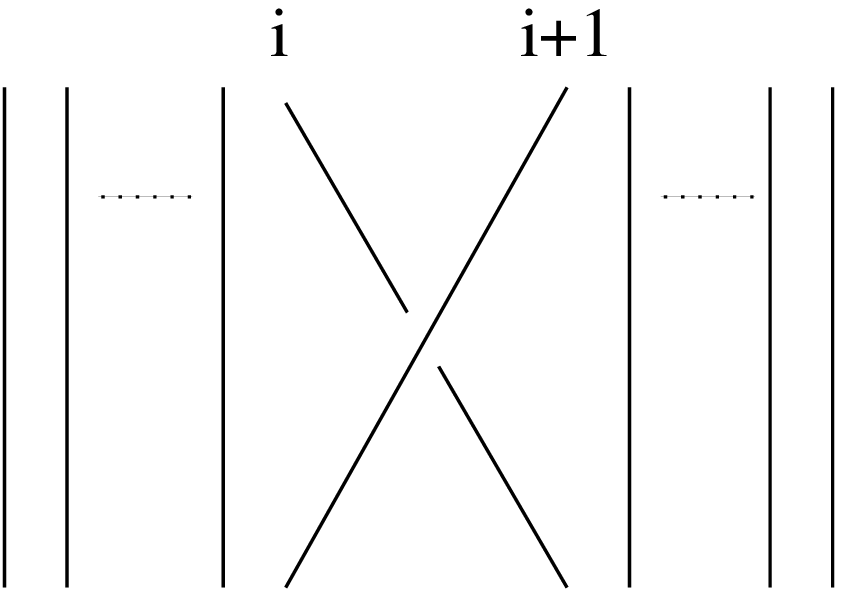}\end{array}.
\ee
Since a bracket state of the closure $\bar b$ of a braid $b$ is obtained by choosing a smoothing for each $\sigma_i^\pm$ in $b$ [according to the relation (\ref{Kau1})], it follows that each state of $\bar b$ corresponds to the strand closure of an element in the Temperley-Lieb algebra. This gives an algorithm to compute $\langle \bar b\rangle$ for any braid $b$ via a sum of trace evaluations  of elements of $TL_n$. More generally, the method can be applied to any tangle, but in the case of the braid group there is a underlying representation $\rho: B_n~\rightarrow~TL_n$ to the Temperley-Lieb algebra, determined by
\be\la{braid-rep}
\rho(\sigma_i)=Ae_i+A^{-1} 1_n\,,~~~~~~\rho(\sigma_i^{-1})=A^{-1}e_i+A 1_n\,,
\ee
from which it follows that
\be
tr(\rho (b)) = \langle \bar b\rangle,
\ee
giving the bracket as a trace on the representation of the braid group into the Temperley-Lieb algebra.

A standard way to apply the Temperley-Lieb recoupling theory to classical $SU(2)$ trivalent spin network evaluations, often called {\it chromatic evaluation}, consists of defining the given trivalent network as a trivalent graph with links labeled by an admissible coloring. More precisely, a link of color $n$ (with $n=2j$, $j$ being the spin Irrep associated with the link) represents $n$ parallel lines and a {\it symmetrizer}, or {\it projector operator} (the reason for this name will be clearer soon). The symmetrizer is defined as 
\be\la{Symm}
\begin{array}{c}\includegraphics[width=1.5cm]{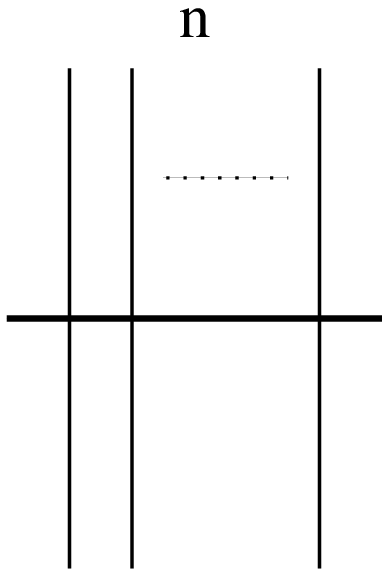}\end{array}=\frac{1}{n!}\sum_p (-1)^{|p|}P^{\va (p)}\,,
\ee
where $P^{\va (p)}$, with $p=1,\dots,n!$, represents the $n$-tangles given by all the possible ways of connecting the $n$ incoming strands with the $n$ outgoing ones, obtained as $n!$ permutations, and $|p|$ is the sign of the permutation.

Since a trivalent spin network is obtained by joining several trivalent vertices by their edges, through this construction, a trivalent spin network determines a closed tangle. One can now use the Kauffman bracket relations (\ref{Kau1}) and (\ref{Kau2}), in the case $A=\pm 1$, for the (chromatic) evaluation of this tangle. 

Contractions of intertwiners and Wigner $3j$-symbols can therefore be computed as chromatic evaluations of colored diagrams, using only the two Penrose identities (for $A=1$)
\ba
&&\left< \begin{array}{c}\psfrag{a}{}\psfrag{b}{}
\includegraphics[width=0.8cm]{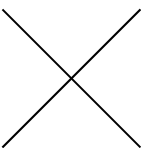}\end{array}\right>=\left<\begin{array}{c}
\includegraphics[width=0.8cm]{qL2.eps}\end{array}\right>+
 \left<\begin{array}{c}
\includegraphics[width=0.8cm]{qL1.eps}\end{array}\right>\la{Pen1}\\
&&\left< \begin{array}{c}\psfrag{a}{}\psfrag{b}{}
\includegraphics[width=0.8cm]{Loop.eps}\end{array}\right>= -2\la{Pen2}\,.
\ea
The expression (\ref{Symm}) for the symmetrizers can be generalized to the case when $q=A^2$ is a $2r$th primitive root of unity, i.e. $q^r=-1, q=\exp{i\pi/r}$. The {\it $q$-symmetrizers} take the form \cite{KL}
\be\la{qSymm}
\begin{array}{c}\includegraphics[width=1.5cm]{Symm1.eps}\end{array}=\frac{1}{\{n\}!}\sum_p (A^{-3})^{|p|}P_q^{\va (p)}\,,
\ee
where $\{n\}!=\prod_{k=1}^n\left(\frac{1-A^{-4k}}{1-A^{-4}}\right)$ is a version of $q$-deformed factorial, which reduces to $n!$ for $A=\pm 1$, and $P_q^{\va (p)}$ still represents the $n$-tangles obtained from all possible permutations of the outgoing strands, but now one has to specify the kind of crossings. In $P_q^{\va (p)}$ all the crossings are of the kind $\begin{array}{c}
\includegraphics[width=0.4cm]{x-quantum.eps}\end{array}$ and they satisfy the first of the relations (\ref{Kau1}). Therefore, all the $P_q^{\va (p)}$ can be written as products of the elementary braids $\sigma_i$ and, by means of the representation (\ref{braid-rep}), they can be expanded to elements in the Temperley-Lieb algebra. Thus, we can regard the $q$-symmetrizers (\ref{qSymm}) as in $TL_n$.

Let us now elucidate why the symmetrizers introduced above are also called {\it projector operators}. For $q$ a $2r$th primitive root of unity, it can be shown that, $\forall ~~n\leq r-1$, there is a unique, non-zero $f_n\in TL_n$ such that
\ba\la{proj}
&&f_n^2=f_n~~~(projection~ operator)\,;\\
&&f_n e_i=e_i f_n=0 ~~~\forall~~i\leq n-1\,. 
\ea
Because of these two properties and its uniqueness, it can be shown that
\be
f_n=\begin{array}{c}\includegraphics[width=1.5cm]{Symm1.eps}\end{array}.
\ee
Therefore, the $q$-symmetrizers are projectors. Since the $f_n$ can be built inductively, this provides an alternative recursive definition for the $q$-symmetrizers (\ref{qSymm}), namely
\ba\la{recursive}
&&\begin{array}{c}\includegraphics[width=0.5cm]{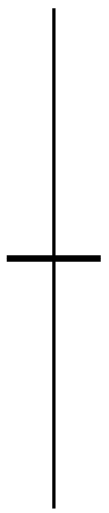}\end{array}=
\begin{array}{c}\includegraphics[width=0.05cm]{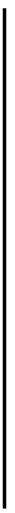}\end{array}\n\\
&&\begin{array}{c}\includegraphics[width=2cm]{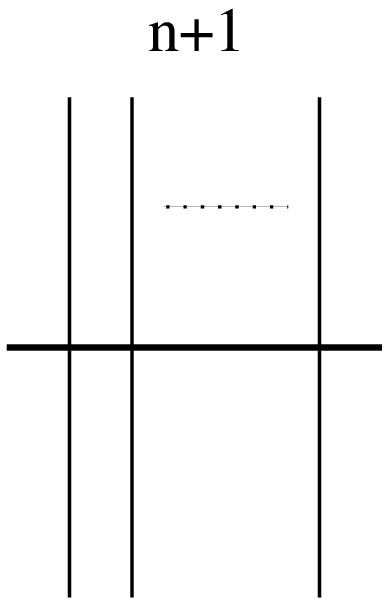}\end{array}=\begin{array}{c}\includegraphics[width=2cm]{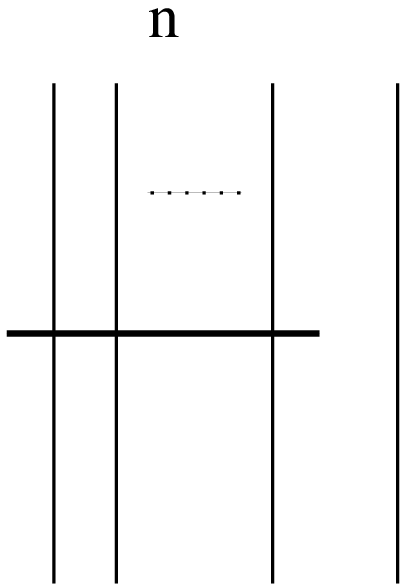}\end{array}-\frac{\Delta_{n-1}}{\Delta_n}\begin{array}{c}\includegraphics[width=3cm]{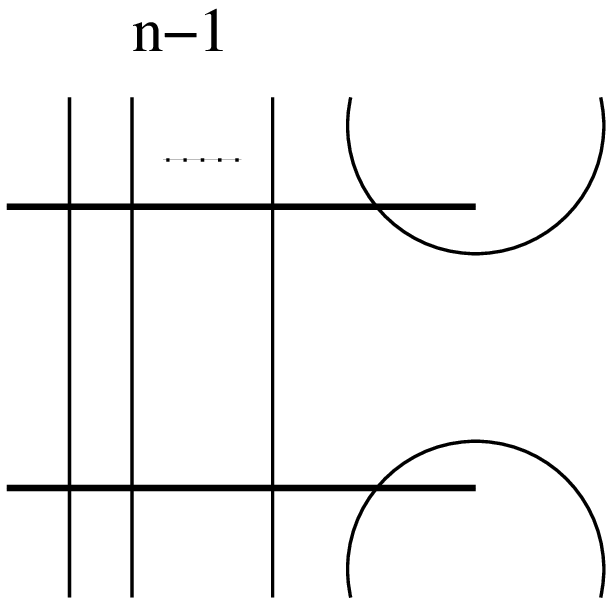}\end{array}\,,
\ea
where
\be\la{Deltan}
\Delta_n=tr( f_n)=\left<\begin{array}{c}\includegraphics[width=3cm]{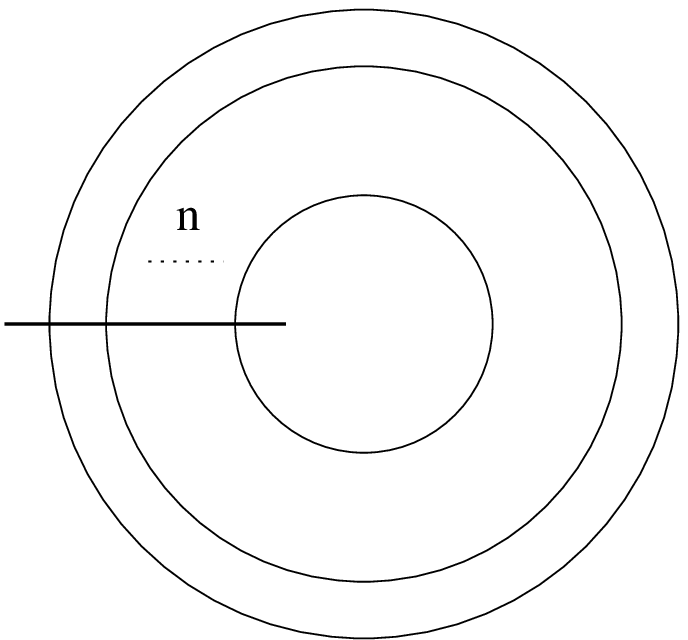}\end{array}\right>=(-1)^n\left(\frac{q^{n+1}-q^{-(n+1)}}{q-q^{-1}}\right)
\ee
are the so-called {\it quantum dimensions}. The last equality in the previous expression can be proven from the recursive definition [obtained from the closure of (\ref{recursive}) and the projectors property (\ref{proj})]
\be\la{recursive2}
\Delta_0=1\,, ~~~~\Delta_1=d\,,~~~~\Delta_{n+1}=d\Delta_n-\Delta_{n-1}\,.
\ee

Let us point out how in both the recursive relations \eqref{recursive} and \eqref{recursive2} the deformation parameter $q=A^2$ does not appear explicitly; in particular, if $\Delta_1$ evaluates to the classical dimension instead, then one recovers the recoupling theory of the `classical' $SU(2)$ group. 


\subsection{Constraint algebra}\la{sec:Algebra}

In order to be able to implement the dynamics correctly, we first need to make sure that the quantum curvature constraint \eqref{Cq} closes the proper algebra. From the classical algebra \eqref{algebra} we see that this amounts to having the action of the commutator of any two operators $\tr\left[N_{p}\,\hat W_{p}\left(A_{\va \Lambda}^{\pm}\right)\right]$ (which belong to the sum over plaquettes providing a regularization for the curvature constraint) to vanish on a gauge invariant state. Given the non-trivial action of the triad field $e_{a}^{i}=\epsilon_{ab}E^{bi}$ (entering the expression of the non-commutative connections $A_{\va \Lambda}^{\pm}$) only on links transversal to the plaquette on which they are smeared, we need to introduce also the  dual complex $\Delta_{\Sigma^*}$ with plaquettes $p^*\in \Delta_{\Sigma^*}$ dual to $p\in \Delta_{\Sigma}$. Then, the set of states to be considered when studying the regularized constraint algebra will be a subset $Cyl(\Delta_{\Sigma^*}) \subset Cyl$ consisting of all cylindrical functions whose underlying graph is contained in the one-skeleton of $\Delta_{\Sigma^*}$.

With this regularization, one can see that the commutator between Wilson loops on different plaquettes gives immediately zero; therefore, we only need to consider the case of two operators defined on the same plaquette. For simplicity, we will consider the action of such a commutator on a gauge invariant state represented by a bivalent node; the calculation can be straightforwardly generalized to higher valent nodes leading to the same implications for its vanishing. We take the state in the fundamental representation $j=1/2$; by means of the recursive relation \eqref{recursive} for the symmetrizers, the calculation can be extended to a generic spin-$j$ representation.  Graphically, the plaquette on which the Wilson loop operator is defined and the state on which it acts are given by
\bee
\begin{array}{c}\includegraphics[width=2.3cm]{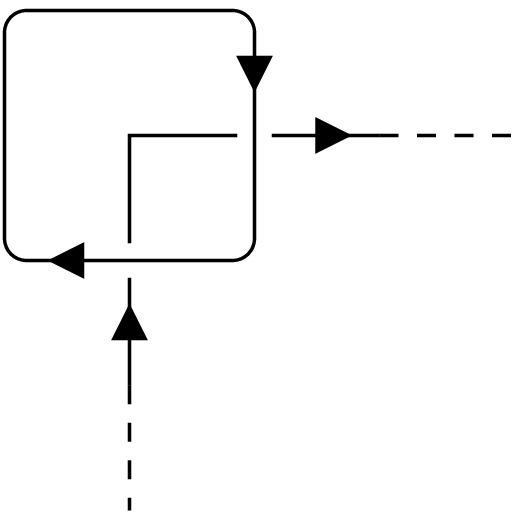}\end{array}\,.
\eee

We can now use the scheme developed in \cite{Anomaly} and the actions \eqref{cross1} and \eqref{cross2} to compute the commutator of the operator $\tr\left[N_{p}\,\hat W_{p}\left(A^{\va \Lambda}_{+}\right)\right]$ with itself (the $A_{-}$ case is analogous), namely
\ba\la{algebra-1/2}
&&\left[\tr\left[N_{p}\,\hat W_{p}\left(A^{\va \Lambda}_{+}\right)\right], \tr\left[M_{p}\,\hat W_{p}\left(A^{\va \Lambda}_{+}\right)\right]\right]\rhd|\lefthalfcap_{\half} \rangle \n\\
&&=\frac{1}{4}\Big[ \begin{array}{c}\includegraphics[width=1.5cm]{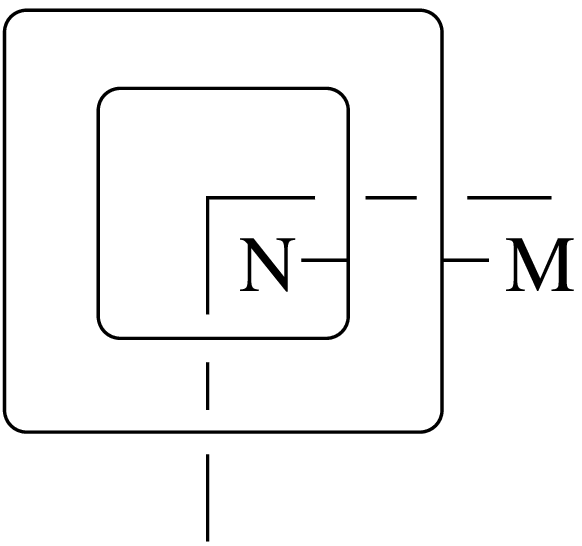}\end{array}+
\begin{array}{c}\includegraphics[width=1.5cm]{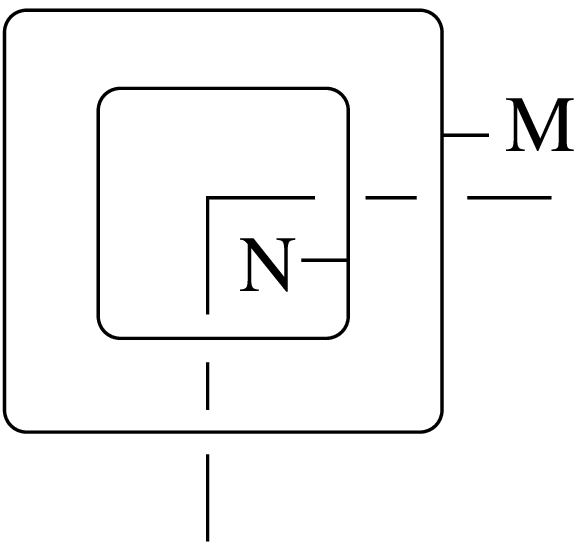}\end{array}+
\begin{array}{c}\includegraphics[width=1.5cm]{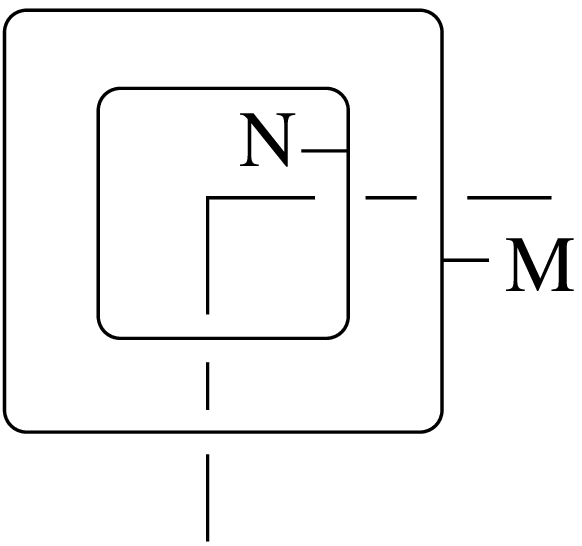}\end{array}+
\begin{array}{c}\includegraphics[width=1.5cm]{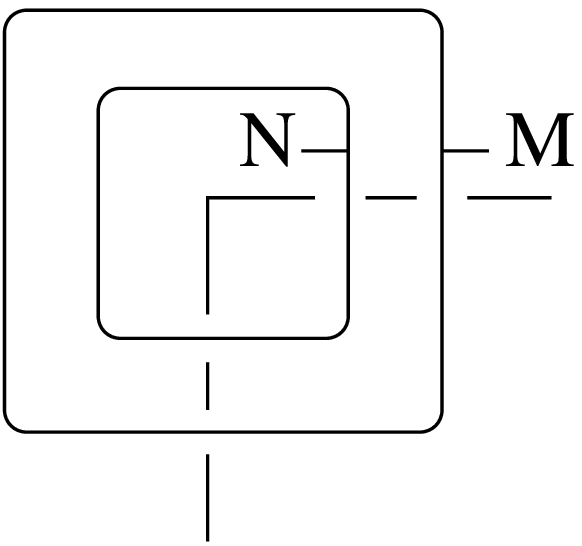}\end{array}- N\leftrightarrow M\Big]\n\\
&&=\frac{1}{4}\Big[\big(A^2+A^{-2}+\!\!\begin{array}{c}\includegraphics[width=.65cm]{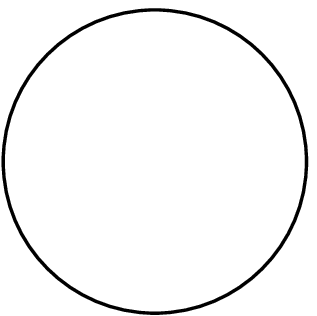}\end{array}\!\!\big)
\big(2\begin{array}{c}\includegraphics[width=1.45cm]{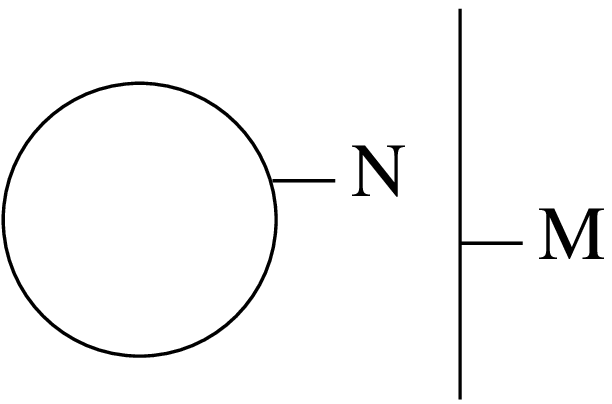}\end{array}
\!\!+\!\!\begin{array}{c}\includegraphics[width=1.15cm]{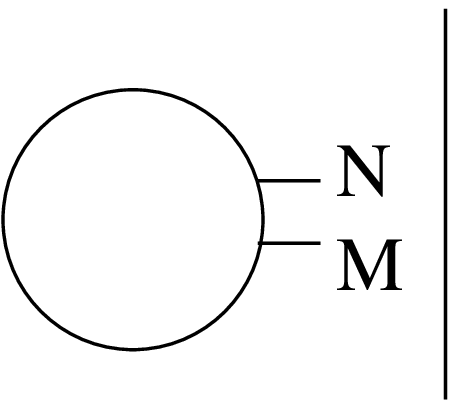}\end{array}
\!\!+\!\!\begin{array}{c}\includegraphics[width=1.15cm]{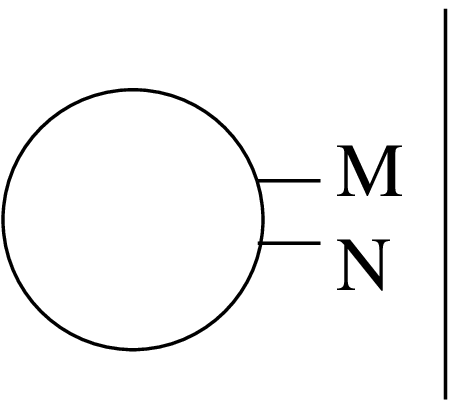}\end{array}
\!\!+\!\!(A^2+A^{-2})(\!\begin{array}{c}\includegraphics[width=.42cm]{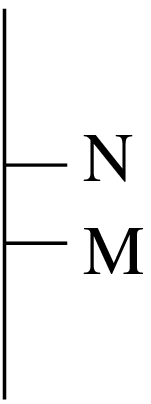}\end{array}\!\!+\!\!
\begin{array}{c}\includegraphics[width=.42cm]{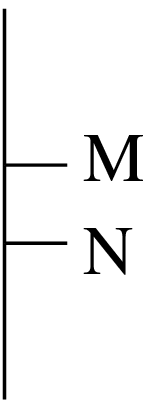}\end{array}\!\!)\big)\n\\
&&+2\begin{array}{c}\includegraphics[width=1.0cm]{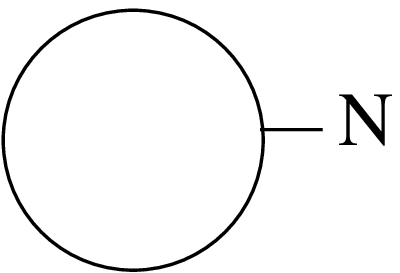}\end{array}
\big(\begin{array}{c}\includegraphics[width=1.05cm]{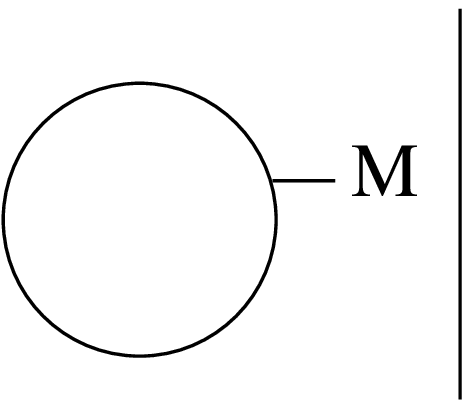}\end{array}
+(A^2+A^{-2}+\!\!\begin{array}{c}\includegraphics[width=.65cm]{Comm5.eps}\end{array}\!\!)\begin{array}{c}\includegraphics[width=.42cm]{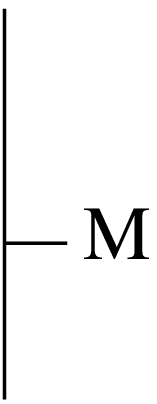}\end{array}\big)- N\leftrightarrow M\Big]\n\\
&&=2\big(A^2+A^{-2}+\!\!\begin{array}{c}\includegraphics[width=.65cm]{Comm5.eps}\end{array}\big)
\big(\begin{array}{c}\includegraphics[width=1.45cm]{Comm7.eps}\end{array}
-\begin{array}{c}\includegraphics[width=1.45cm]{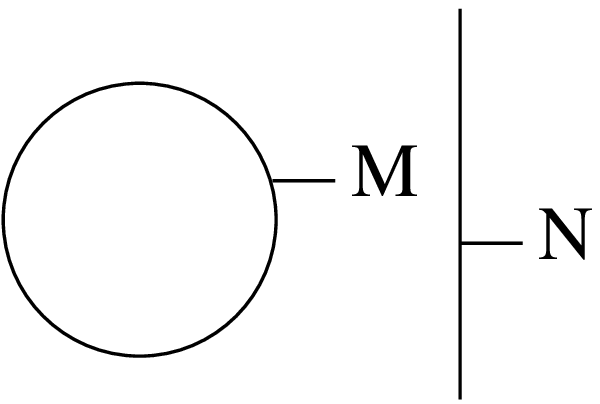}\end{array}\big)\,.
\ea
A similar calculation for the state in a generic spin-$j$ representation leads to
\be\la{algebra-j}
\left[\tr\left[N_{p}\,\hat W_{p}\left(A^{\va \Lambda}_{+}\right)\right], \tr\left[M_{p}\,\hat W_{p}\left(A^{\va \Lambda}_{+}\right)\right]\right]\rhd|\lefthalfcap_{j} \rangle =
2\big(A^2+A^{-2}+\!\!\begin{array}{c}\includegraphics[width=.65cm]{Comm5.eps}\end{array}\!\big)
\big(\!\begin{array}{c}\includegraphics[width=1.45cm]{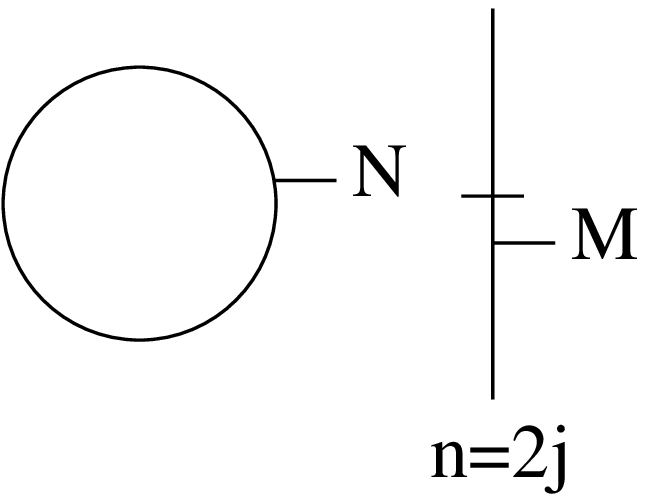}\end{array}
-\begin{array}{c}\includegraphics[width=1.45cm]{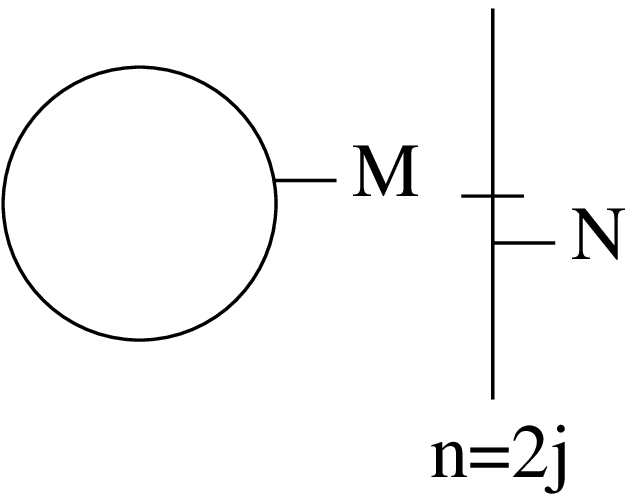}\end{array}\!\big)\,.
\ee

Hence we see that the action of the commutator on a gauge invariant state does not vanish unless the infinitesimal loop evaluates to the spin-$1/2$ quantum dimension, namely 
\be\la{q-dim}
\begin{array}{c}\includegraphics[width=1.cm]{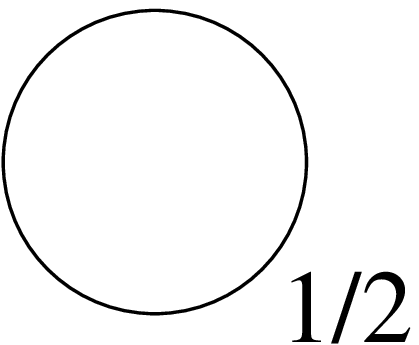}\end{array}=-(A^2+A^{-2})\,.
\ee
Notice that such condition is also necessary for the Reidemeister move of type II
\be
\begin{array}{c}\includegraphics[width=2.5cm]{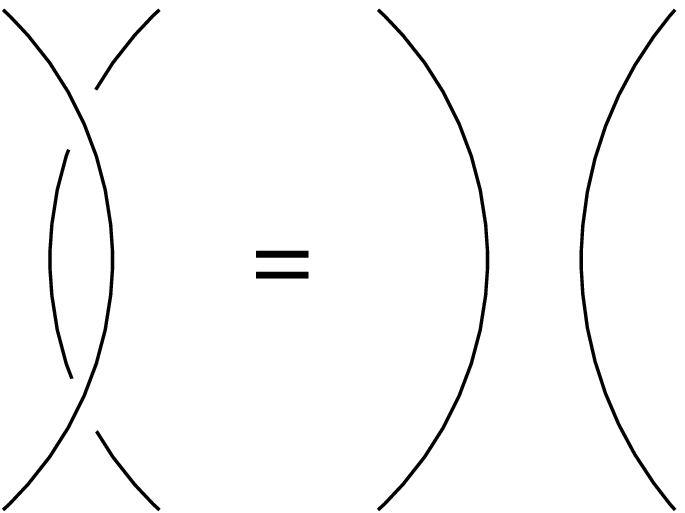}\end{array}
\ee
to hold; the validity of this topological relation is important to eliminate ambiguities in the definition of the physical scalar product introduced in the next section.

Furthermore, if we plug \eqref{q-dim} in the recursion relation \eqref{recursive2}, it can be shown that an infinitesimal loop in a generic spin-$j$ representation evaluates then to the spin-$j$ quantum dimension \eqref{Deltan}, namely\footnote{From the relation $\sqrt{\Lambda}=2\pi/k$, obtained by rewriting (two copies of) the Chern-Simons action with level $k$ in terms of the connection and triad variables of first order gravity with cosmological constant, one recovers the expression \eqref{Deltan} for the $\su(2)_q$ quantum dimension, where the integer $r$ corresponds to the level $k$ (up to a shift of 2).}
\be\la{q-dim-j}
\begin{array}{c}\includegraphics[width=.85cm]{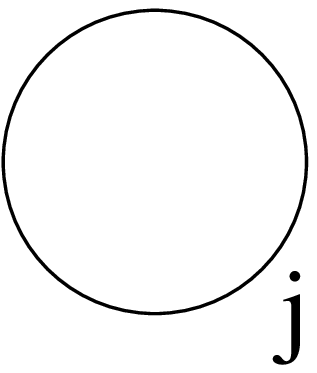}\end{array}=(-)^{2j}[2j+1]_q=(-)^{2j}\frac{q^{2j+1}-q^{-(2j+1)}}{q-q^{-1}}\,.
\ee


An immediate consequence of the evaluation of the infinitesimal loop to the quantum dimension is that the skein relation
\be\label{integration-old}
\begin{array}{c}  \includegraphics[width=1.2cm,angle=360]{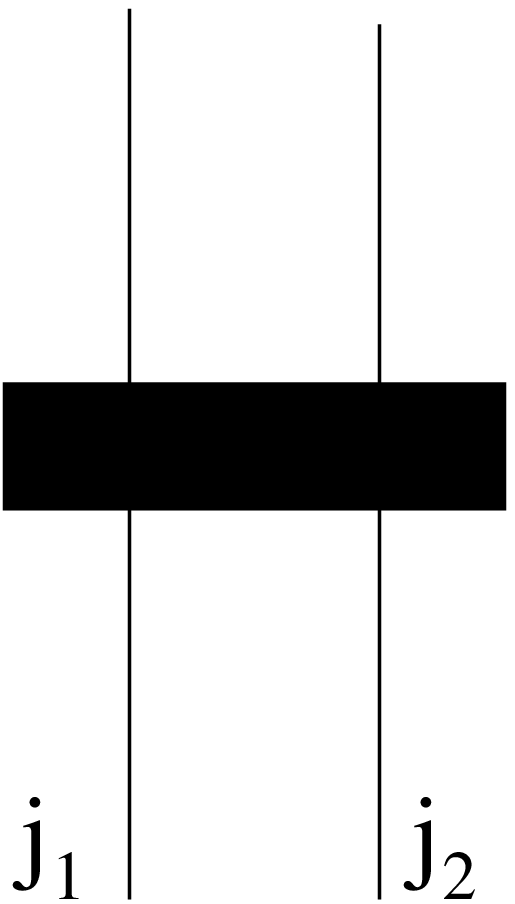}\end{array}=\frac{1}{[2j_1+1]}\delta_{j_1 j_2}\begin{array}{c}  \includegraphics[width=1cm,angle=360]{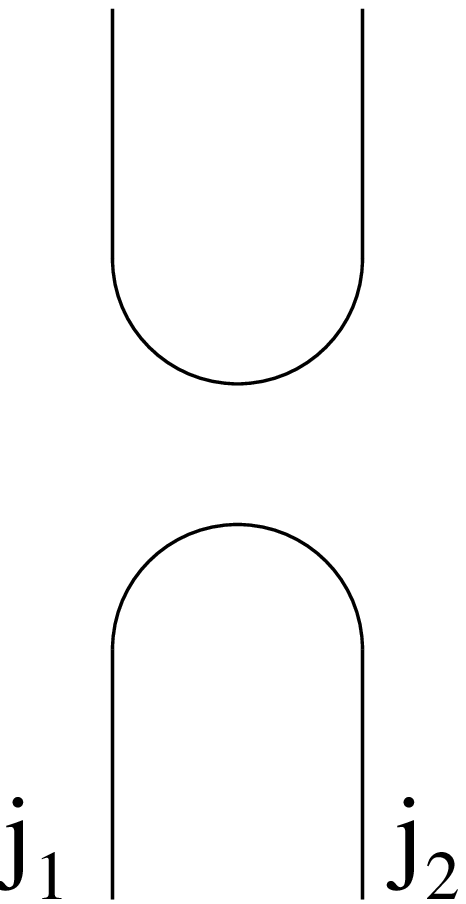}\end{array}
\ee
encoding the integration over a group element 
in common between the two edges (symbolized by a black box) has to be modified. The reason for this modification comes from the fact that, while taking into account the condition \eqref{q-dim-j} for an anomaly-free algebra of constraints, we do not want to change the properties of the Ashtekar-Lewandowski measure \cite{AL}. In particular, if we indicate by an index $q$ the box integration corresponding to the  modified skein relation, we still want the projector property to hold \cite{Oeckl}, namely
\be\label{projector}
\begin{array}{c}  \includegraphics[width=1.3cm,angle=360]{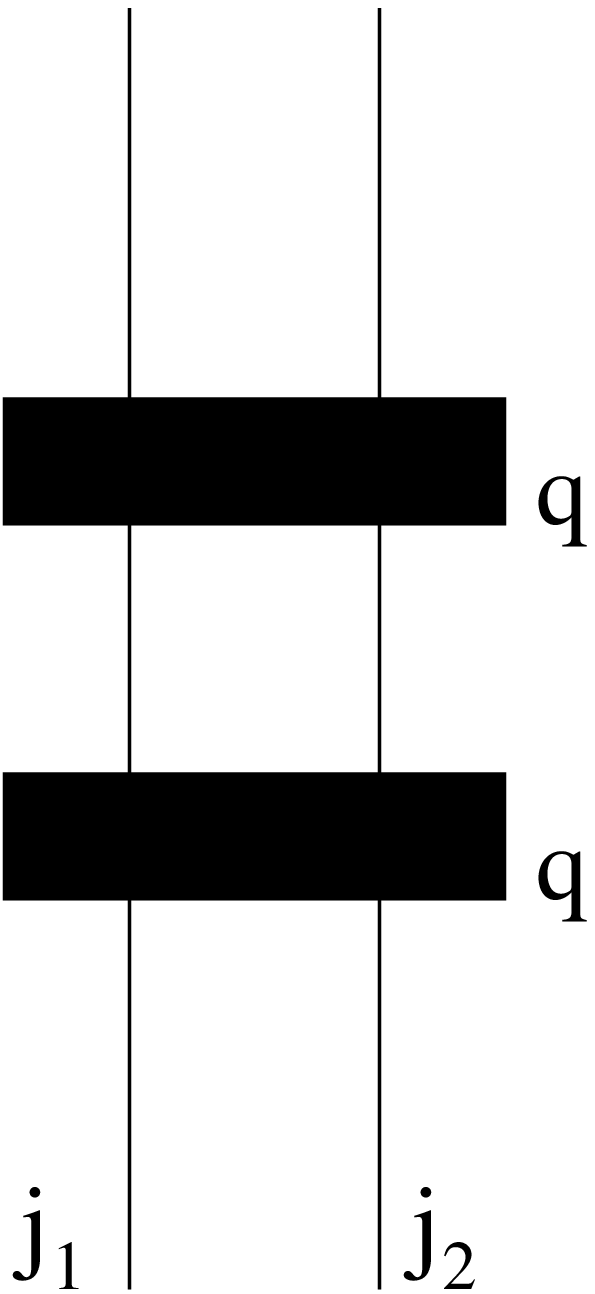}\end{array}=\begin{array}{c}  \includegraphics[width=1.3cm,angle=360]{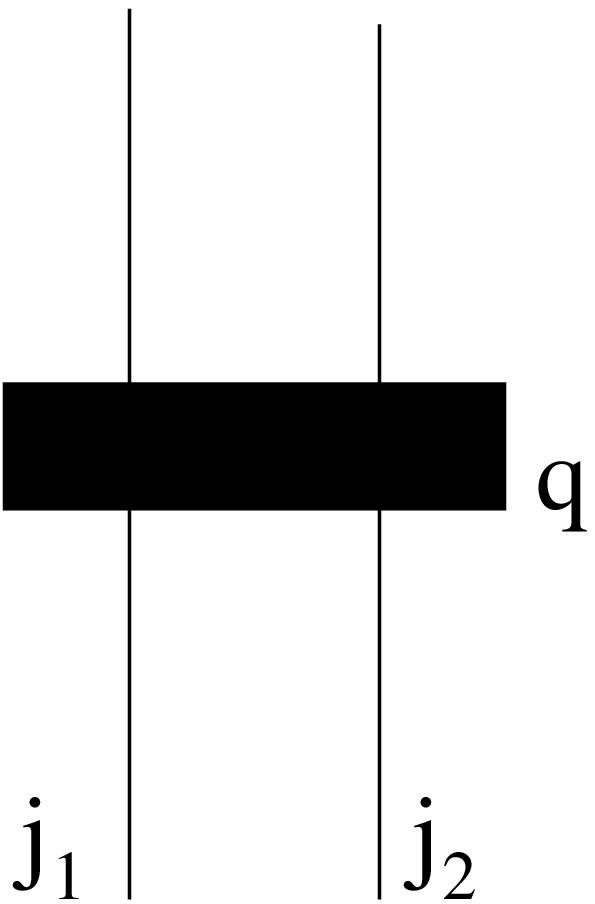}\,.\end{array}
\ee
For this to be the case, it is enough to renormalize the box integration as
\be
\begin{array}{c}  \includegraphics[width=1.3cm,angle=360]{projector-r}\end{array}=C \begin{array}{c}  \includegraphics[width=1.15cm,angle=360]{ggl}\,,\end{array}
\ee
where the factor $C$ has to be determined by imposition of the property \eqref{projector}. By applying \eqref{integration-old} twice, we get
\ba
\begin{array}{c}  \includegraphics[width=1.15cm,angle=360]{projector-l}\end{array}&=&C^2\begin{array}{c}  \includegraphics[width=1cm,angle=360]{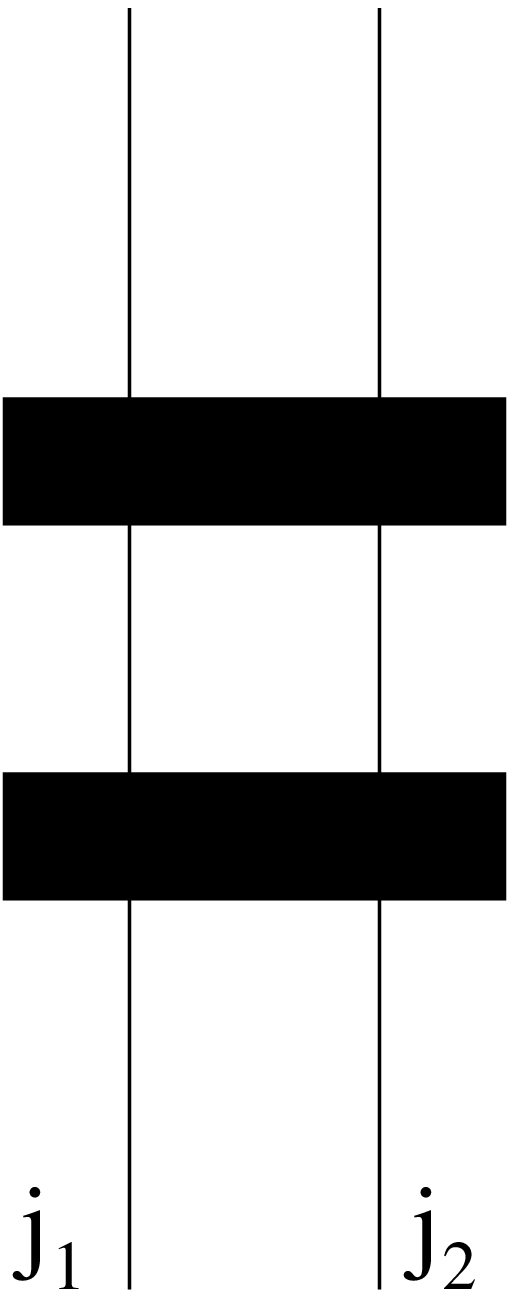}\end{array}= \frac{C^2}{[2j_1+1]}\delta_{j_1 j_2}\begin{array}{c}  \includegraphics[width=1cm,angle=360]{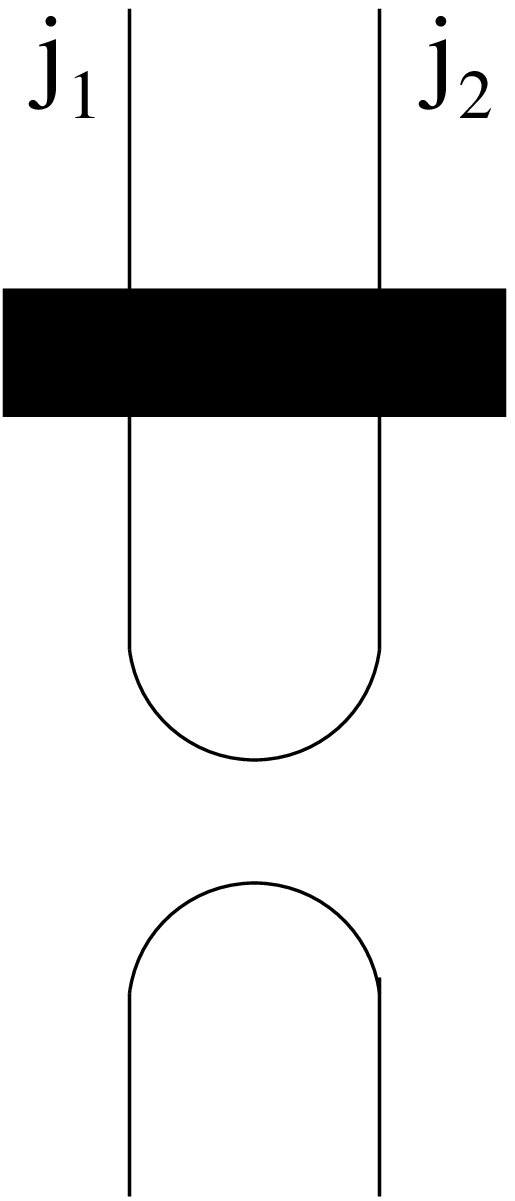}\end{array}=
\frac{C^2}{[2j_1+1]^2}\delta_{j_1 j_2}\begin{array}{c}  \includegraphics[width=1cm,angle=360]{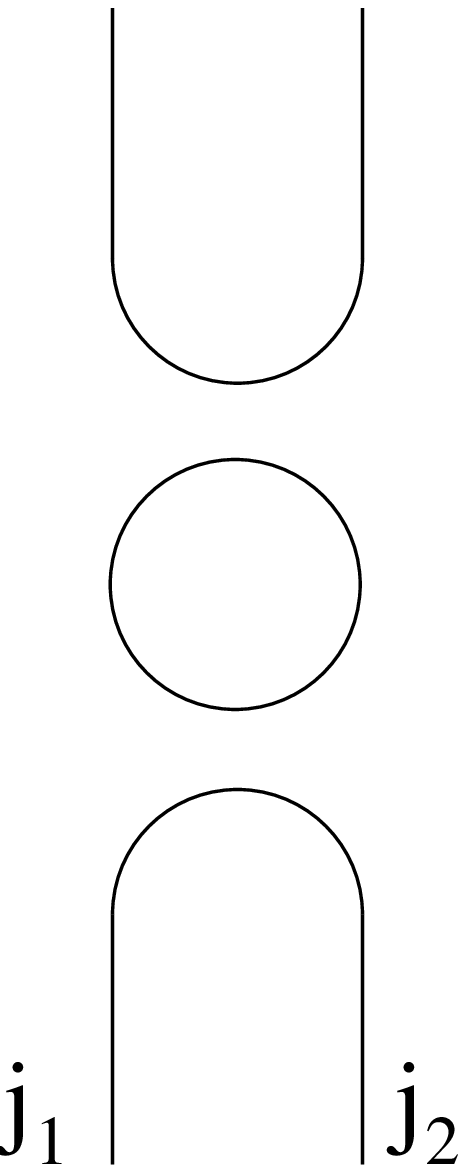}\end{array}\n\\
&=&C \frac{[2j_1+1]_q}{[2j_1+1]} \Big(C \begin{array}{c}  \includegraphics[width=1cm,angle=360]{ggl}\end{array}\Big)=
C \frac{[2j_1+1]_q}{[2j_1+1]}\begin{array}{c}  \includegraphics[width=1.15cm,angle=360]{projector-r}\end{array}
\ea
and, therefore, the box integration map satisfies the projector property \eqref{projector} if and only if
\be
C=\frac{[2j_1+1]}{[2j_1+1]_q}\,.
\ee
This implies that the skein relation \eqref{integration-old}, which is no longer compatible with the  Ashtekar-Lewandowski measure properties,  has to be replaced by
\be\label{integration}
\begin{array}{c}  \includegraphics[width=1.3cm,angle=360]{projector-r}\end{array}=\frac{1}{[2j_1+1]_q}\delta_{j_1 j_2}\begin{array}{c}  \includegraphics[width=1cm,angle=360]{ggr}\,.\end{array}
\ee
We are shortly going to use this relation to compute physical amplitudes of the canonical quantum theory.

\subsection{Physical Transition Amplitudes}\la{sec:Projector}

We are ready to introduce the projector operator defining the physical scalar product of 2+1 LQG with positive cosmological constant. As a consequence of the non-commutativity of the connection \eqref{conn} and the following crossing relations \eqref{cross1}, \eqref{cross2} of its holonomy, we saw in the previous section that the algebra of the quantum curvature constraint \eqref{Cq} does not close unless the infinitesimal loop evaluates to the quantum dimension. This leads us to define the analog of the scalar product \eqref{scalar} in the case of $\Lambda>0$ as
\ba\label{scalarq}
 \langle s, \;s^{\prime}\rangle_{\va ph-\Lambda}=\langle P^{\va \Lambda} s,\;s^{\prime} \rangle &:=& \lim_{\epsilon\rightarrow 0} \ \ \langle\prod_{p} \
{\delta}(W^{\va \Lambda}_{p})s, \; s^{\prime}\rangle \n\\
&=& \lim_{\epsilon\rightarrow 0} \ \  \sum_{j_{\va p}}\
 [2j_{\vani p}+1]_q \langle \prod_{p}
 \chi_{\va j_p}(W^{\va \Lambda}_{p})\ s,\; s^{\prime}\rangle\,,
 \ea
where $W^{\va \Lambda}_{p}\equiv W_{p}(A^{\va \Lambda})$. The replacement of the classical dimension with the quantum dimension in the projector operator follows automatically from the loop redefinition \eqref{q-dim-j} and the renormalization \eqref{integration}, which morally correspond to replacing the classical $SU(2)$ recoupling theory with the quantum group one. However, such a replacement is not done by hand but is forced on us by the non-trivial action of the (quantum version of the) curvature constraint \eqref{Cq} in presence of a non-vanishing cosmological constant, as reflected in the constraint algebra \eqref{algebra-j}.\footnote{A way to see how the quantum dimension appears in the delta function expansion, entering the physical projector, is to use it to express the requirement [from \eqref{algebra-j}] of the infinitesimal loop to evaluate to the quantum dimension, explicitly
\be
[2j+1]_q=\int dg \chi_j(g)\delta(g^{\va \Lambda})=\sum_{j'}\alpha_{j'}\int dg \chi_j(g) \chi_{j'}(g)=\alpha_j\delta_{jj'}\,.
\ee
} Moreover, due to the modified relations \eqref{q-dim-j} and \eqref{integration}, encoding the $U_q(su(2))$ recoupling theory, the presence of the quantum dimension in the expression \eqref{scalarq} is crucial in order to preserve the topological invariance of the physical scalar product.

Let us show that the expectation values of spin network observables defined by \eqref{scalarq} reproduce the Turaev-Viro model amplitudes. In order to compute the physical inner product between classical kinematical spin network states, we do not want to introduce the deformation parameter $q=A^2$ in the initial and final states. This can be done by replacing each link in  $s, s'$ with a combination of strands and loops given by the relation \eqref{recursive}; by correctly joining all the strands at each intertwiner, the two closed spin network graphs associated to the states  $s, s'$ can in this way be expressed as a combination of products of loops, with no appearance of powers of $A$ factors. So far we cannot yet distinguish between classical and quantum group chromatic evaluation. To recover the bracket polynomial \eqref{bracket} then we simply need to show that the physical transition amplitude between products of loops is equal to the products of the quantum dimensions in the spin-$j$ representations coloring  the respective loops. We are now going to prove that this is indeed the case.

We take $s$ to be the vacuum state and $s'$ a collection of $N$ loops with associated Irreps labeled by the spins $k_1,\dots k_N$---we could similarly take $s$ to be a collection of $N/2$ loops and $s'$ a collection of the other half or any other subdivision and the final result would not change. We assume $\Sigma$ to have the topology of a sphere, i.e. Euler characteristic $\chi=2$. As in the case of $\Lambda=0$, a redundancy then appears in the product of delta distributions in the expression of the projection operator; this is a consequence of the discrete analog of the Bianchi identity and the correct result can be obtained by eliminating a single arbitrary plaquette holonomy $U_p$ from the product. The $N$ loops could be taken either all concentric or all disjoint from each other or a combination of the two; again, the final result is left unchanged. In this example we are going to consider the $N$ loops all inside of each other and we are going to remove a plaquette holonomy from the projector operator in \eqref{scalarq} that is outside of all the loops. With these choices, the physical scalar product reads
\be\la{scal1}
< P^{\va \Lambda} \emptyset,\!\!\begin{array}{c}  \includegraphics[width=1.3cm]{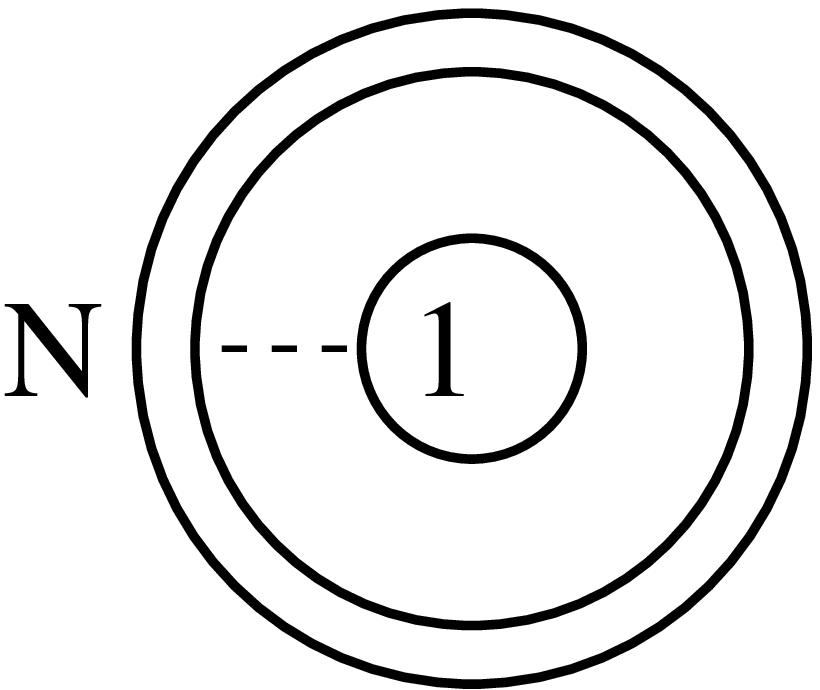}\end{array}\!\! >~=~
\lim_{\epsilon\rightarrow 0}\int  \left(\prod_h dg_{\va h}\right) \prod_n\chi_{\va k_n}(g_{\va n})\prod_p \sum_{j_p}[2j_p+1]_q~\chi_{\va j_p}(W^{\va \Lambda}_{p})\,,
\ee
where $g_{\va n}$ is the holonomy around the $n$th loop in $s'$ and $dg_{\va h}$ corresponds to the invariant $SU(2)$-Haar measure satisfying the skein relation \eqref{integration}. The graph in the scalar product above can be graphically represented as
\begin{figure*}[h!]
\centering
  \includegraphics[width=5.cm]{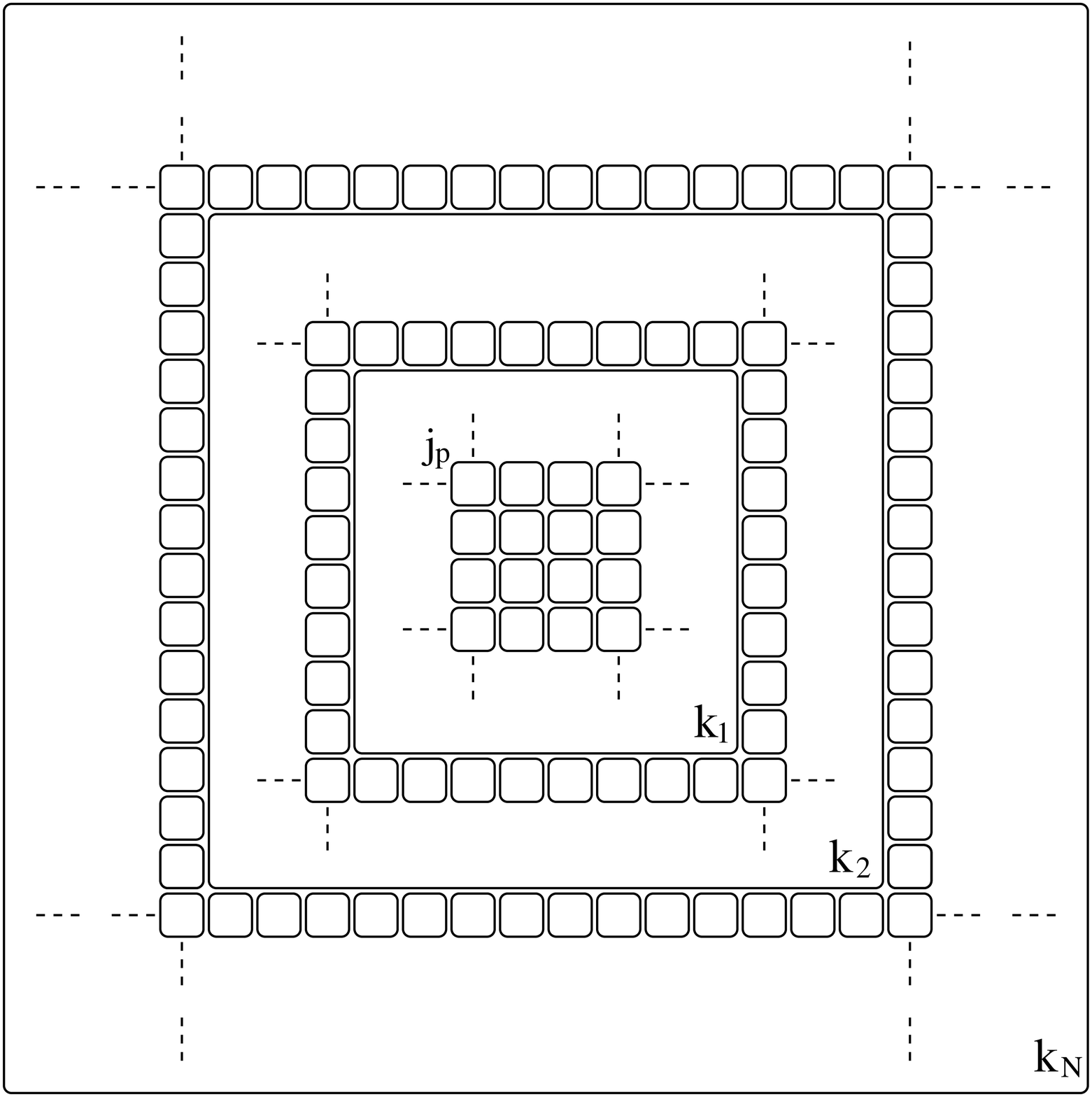}\,.
\end{figure*}

By means of the relation \eqref{integration} we can first integrate over the edges belonging to the plaquettes between any two loops of the state $s'$; by doing so we fuse all the plaquettes between the two loops into only two new loops: one adjacent to the external loop in $s'$ and one adjacent to the internal loop in $s'$. None of these new loops carry a quantum dimension anymore, since all the regions integrated over have the topology of a torus except the new loop inside $k_{\va1}$ which has the topology of a disk. Using a graphical notation, we can then write the scalar product \eqref{scal1} as
\ba\la{scal1}
\langle P^{\va \Lambda} \emptyset,\!\!\begin{array}{c}  \includegraphics[width=1.3cm]{scalar1.eps}\end{array}\!\! \rangle&=&
\prod_{n=1}^N \sum_{j_n} [2j_1+1]_q \begin{array}{c}  \includegraphics[width=4cm]{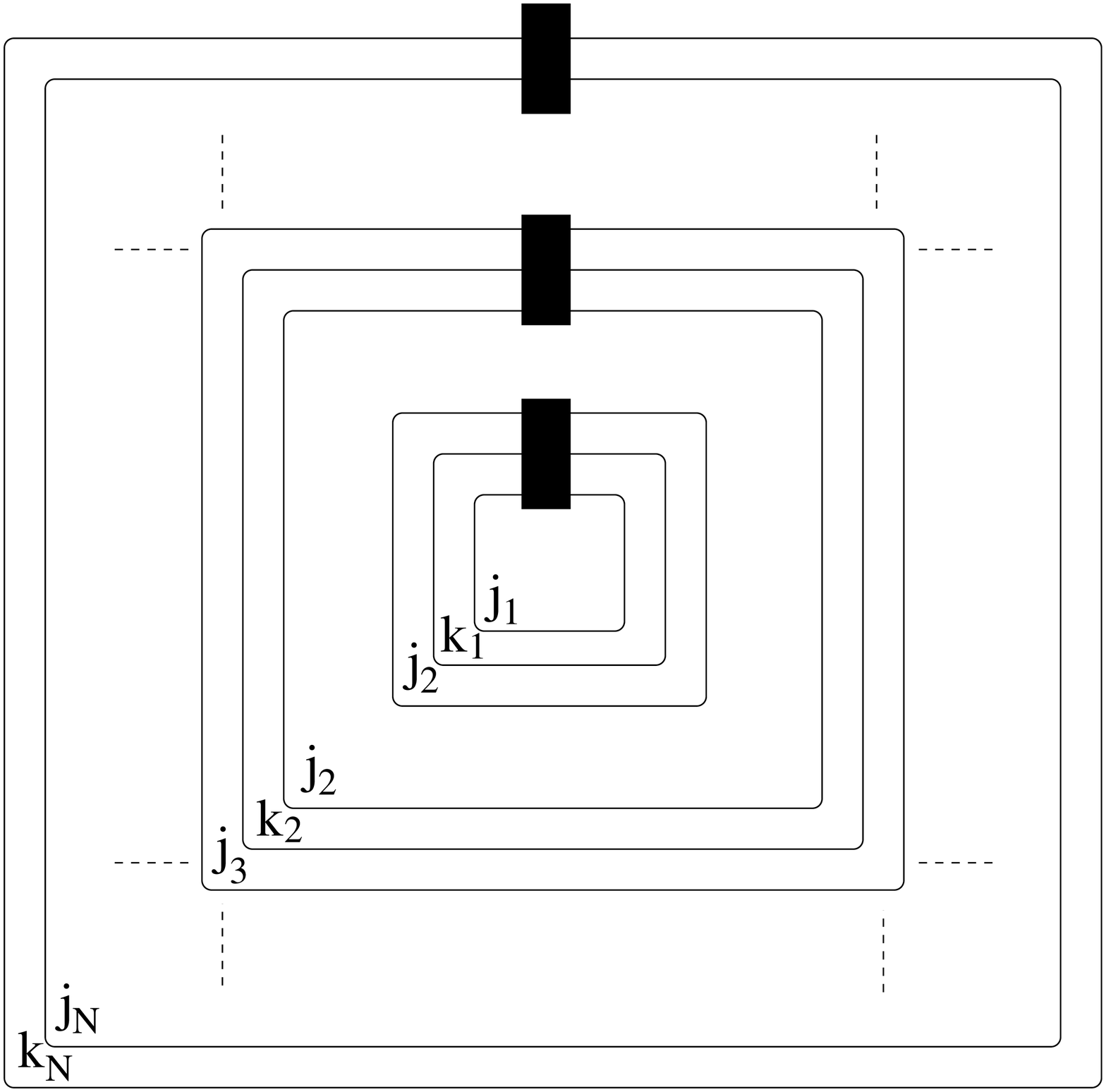}\end{array}\n\\
&=&\prod_{n=2}^N \sum_{j_n}\sum_{j_1=|k_1-j_2|}^{k_1+j_2}[2j_1+1]_q\begin{array}{c}  \includegraphics[width=2.5cm]{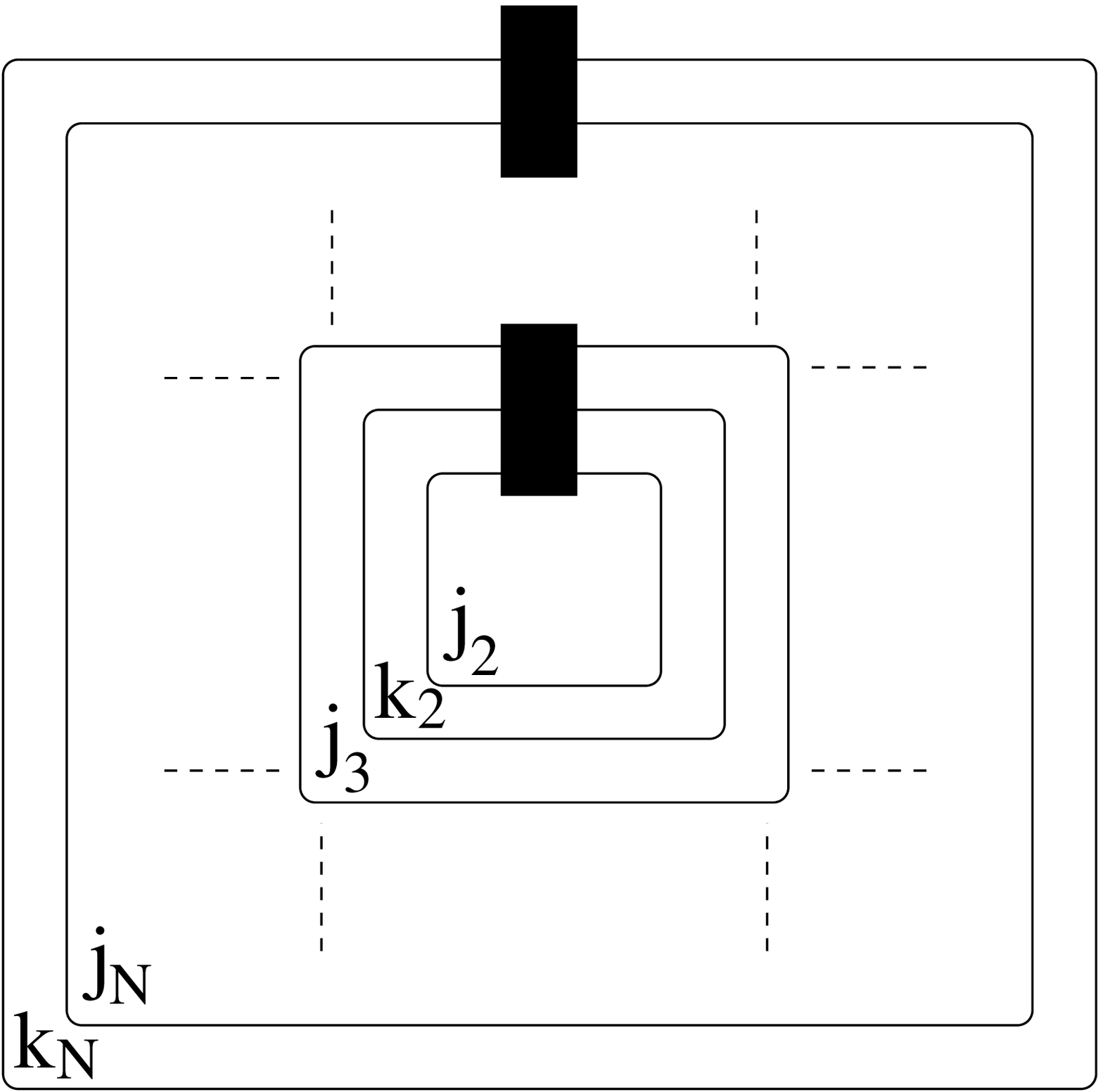}\end{array} \n\\
&=&[2k_1+1]_q \prod_{n=2}^N \sum_{j_n} [2j_2+1]_q\begin{array}{c}  \includegraphics[width=2.5cm]{scalar4.eps}\end{array}\n\\ 
&=&\cdots~=\prod_{n=1}^{N-1}[2k_n+1]_q \sum_{j_N} [2j_N+1]_q \begin{array}{c}  \includegraphics[width=1.1cm]{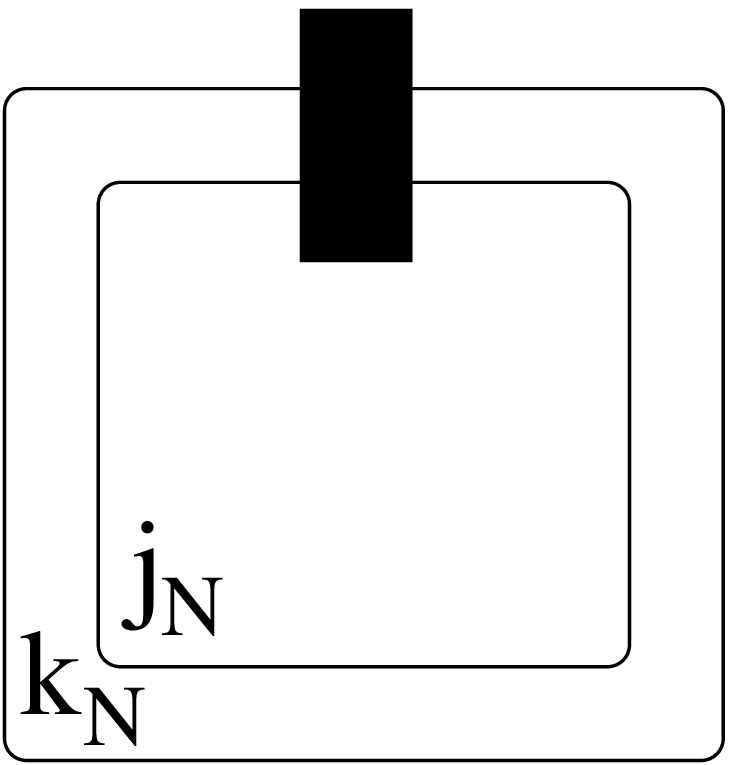}\end{array}\n\\&=&\prod_{n=1}^N[2k_n+1]_q\,.
\ea

This concludes the proof of the equivalence between the canonical loop quantization and the Turaev-Viro spin foam quantization of 2+1 Riemannian gravity with positive cosmological constant.

\section{CONCLUSIONS}\la{sec:Conclusions}

Three-dimensional Riemannian quantum gravity represents a fairly well understood toy model, which constitutes an important consistency check of any candidate theory of quantum gravity in higher dimensions. The theory can be written in two equivalent formulations at the classical level, the BF and Chern-Simons ones. Each formulation can  be then quantized either in the canonical or in the covariant approach. 
On the Chern-Simons side, in both the $\Lambda=0$ and $\Lambda>0$ cases, the two schemes provide mutually consistent results. On the BF side, this program has been completed so far only in the $\Lambda=0$ case, where
the spin foam representation of the path integral has been obtained from the physical scalar product between states of the kinematical Hilbert space. 

Progress towards the canonical quantization of the positive cosmological constant case has been previously made by quantizing the holonomy of a non-commutative connection, in terms of which the dynamics of the theory can be rewritten. In this work, we built on this result to study first the algebra of the curvature constraint. We showed how the quantum dimension emerges inevitably, in order to have an anomaly-free algebra and hence be able to proceed with the quantum imposition of the dynamics. As an immediate consequence, preservation of the Ashtekar-Lewandowski measure properties, in light of the quantum dimension evaluation of an infinitesimal loop, implies a modification of the skein relation associated to the box integration. 

Effectively, the redefinition of the loop \eqref{q-dim-j} together with the renormalization of the box \eqref{integration} amounts to replacing the classical $SU(2)$ recoupling theory by the quantum group one. However, in our approach the Lie algebra deformation is brought in by the dynamics. This can be seen from the central role played by the Kauffman bracket (i) structure for the crossing of the quantum non-commutative holonomies, in terms of which the curvature constraint is expressed. The actions \eqref{cross1} and \eqref{cross2} represent the quantum group seed, from which the algebra \eqref{algebra-j} is derived and the rest of the quantum group structure follows.
In this sense, our analysis differs from the proposals of \cite{Major-Smolin, Dupuis-Girelli} to include a cosmological constant in three-dimensional LQG, where the quantum group is introduced by hand from the beginning, already at the kinematical level. 

With these results in hand then the expression \eqref{scalarq} for the physical scalar product of the theory is a straightforward generalization of the $\Lambda=0$ case \cite{Noui-Perez} and we have shown how this allows us to recover the Turaev-Viro state-sum amplitudes. Therefore, our analysis finally closes the proof of the long conjectured equivalence between canonical and covariant quantization of three-dimensional gravity both in the Chern-Simons and BF formulations.
This represents a highly nontrivial test for the loop approach to quantum gravity, showing complete agreement with other well defined quantization schemes. Furthermore, the present work puts on more solid ground the proposal \cite{Han} to include a cosmological constant in four-dimensional spin foam models by replacing the Lorentz group representations with their quantum group analog.

 Let us point out that the focus on a positive cosmological constant case was motivated by making contact with the Turaev-Viro state sum model \cite{TV} and Witten's treatment \cite{Witten}. However, there is no required restriction on the sign of $\Lambda$ for our analysis to go through and, in case of a negative cosmological constant, the physical scalar product \eqref{scalarq} is still well defined and generates transition amplitudes of the Turaev-Viro model with $q\in \R$.  The avoidance of problems related to the non-compactness of the symmetry group in case of a negative $\Lambda$ is again related to the fact that the kinematical framework of the theory is still the classical $SU(2)$ one and symmetry modifications are introduced only at the dynamical level. Extension of our approach to the Lorentzian case though would seem less immediate, in particular in the $\Lambda<0$ case, where a well defined quantization procedure is not known in any approach.

The equivalence between the Chern-Simons combinatorial quantization and canonical LQG shown above could be used to provide a definition of $SU(2)$ isolated horizons \cite{IH, Review} completely within the loop formalism, along the lines of the program started in \cite{Sahlmann}. This is left for future investigation.

\section*{Acknowledgements}

I am very thankful to Alejandro Perez for several discussions on this topic over the years and his comments on a draft version of this manuscript.

\end{document}